\documentclass[reprint,
 amsmath,amssymb,
 aps,
]{revtex4-1}

\usepackage{graphicx}% Include figure files
\usepackage{dcolumn}% Align table columns on decimal point
\usepackage{bm}% bold math
\usepackage{xcolor}
\begin{document}

\preprint{APS/123-QED}

\title{Veiled Talbot effect}

\author{Murat Yessenov$^{1}$}
\thanks{These authors contributed equally to this work}
\author{Layton A. Hall$^{1}$}
\thanks{These authors contributed equally to this work}
\author{Sergey A. Ponomarenko$^{2,3}$}
\author{Ayman F. Abouraddy$^{1}$}
%\email{raddy@creol.ucf.edu}
\affiliation{$^{1}$CREOL, The College of Optics \& Photonics, University of Central~Florida, Orlando, FL 32816, USA\\$^{2}$Department of Elect. and Computer Eng., Dalhousie University, Halifax, Nova Scotia B3J 2X4, Canada\\$^{3}$Department of Physics and Atmospheric Science, Dalhousie University, Halifax, Nova Scotia B3H 4R2, Canada}

\begin{abstract}
A freely propagating optical field having a periodic transverse spatial profile undergoes periodic axial revivals -- a well-known phenomenon known as the Talbot effect or self-imaging. We show here that introducing tight spatio-temporal spectral correlations into an ultrafast pulsed optical field with a periodic transverse spatial profile eliminates all axial dynamics in physical space while revealing a novel space-time Talbot effect that can be observed only when carrying out time-resolved measurements. Indeed, 'time-diffraction' is observed whereupon the temporal profile of the field envelope at a fixed axial plane corresponds to a segment of the spatial propagation profile of a monochromatic field sharing the initial spatial profile and observed at the same axial plane. Time-averaging, which is intrinsic to observing the intensity, altogether veils this effect.
\end{abstract}

\maketitle

%\section{Introduction}

The long-known Talbot effect \cite{Talbot36PM,Berry96JMO}, or self-imaging \cite{Montgomery67JOSA}, refers to a freely propagating paraxial optical field (at a wavelength $\lambda_{\mathrm{o}}$) having a \textit{periodic} transverse spatial profile (of period $L$) undergoing periodic axial revivals (at planes separated by the Talbot distance $z_{\mathrm{T}}\!=\!\tfrac{2L^{2}}{\lambda_{\mathrm{o}}}$). The field evolution unique to the Talbot effect can be readily observed in physical space. Space-time duality \cite{Kolner94IEEEJQE} suggests that an analogous temporal Talbot effect \cite{Jannson81JOSA,Mitschke98OPN} occurs when a periodic train of pulses (of period $T$) travels in a dispersive medium (with dispersion parameter $k_{2}$): the pulses initially disperse but are subsequently reconstituted axially at multiples of $z_{\mathrm{T}}\!=\!\tfrac{T^{2}}{\pi k_{2}}$. This temporal evolution can be observed by a sufficiently fast detector. The Talbot effect has been used in a wide span of applications \cite{Wen13AOP} extending most recently from structured illumination in fluorescence microscopy \cite{Chowdhury18arxiv} to temporal cloaking \cite{Li17OL} and prime-number decomposition \cite{Pelka18OE}.

An implicit -- yet fundamental -- assumption underlying the Talbot effect is the separability of the spatial and temporal degrees of freedom of the optical field. Imposing a periodic profile in either space or time implies discretizing the corresponding spectrum at multiples of $\tfrac{2\pi}{L}$ or $\tfrac{2\pi}{T}$, respectively, which raises a question with regards to the Talbot effect for pulsed optical fields in which the spatial and temporal degrees of freedom are inextricably intertwined. Specifically, the spectral support domain of so-called `space-time' (ST) wave packets tightly associates each \textit{spatial} frequency with a single \textit{temporal} frequency (or wavelength) \cite{Donnelly93PRSLA,Kondakci19OL,Yessenov19PRA}, such that discretization of one degree of freedom inescapably entails discretization of the other. To date, such pulsed beams \cite{Turunen10PO,FigueroaBook14} have been synthesized exclusively with continuous spectra tailored to render them propagation invariant \cite{Saari97PRL,Kondakci17NP,Wong17ACSP2}. Indeed, it is now well-established that they are transported rigidly in free space at a fixed, but arbitrary group velocity \cite{Salo01JOA,Kondakci19NC,Bhaduri19Optica}. Self-imaging of ST wave packets lacking themselves a periodic structure (thus retaining a continuous spectrum) was examined theoretically by superposing wave packets of different phase velocities \cite{Reivelt02OE}, and realized in space by superposing pulsed Bessel beams \cite{Bock17OL} (self-imaging in both cases is thus unrelated to the Talbot effect).

We pose here a question regarding the propagation of ST wave packets having periodic transverse spatial profiles, whereupon the spatial \textit{and} temporal spectra are simultaneously discretized on account of their tight association. Two mutually exclusive scenarios appear to be on offer. If the spatio-temporal structure undergirding propagation-invariance plays the same role with discrete spectra as with continuous spectra, then any axial dynamics will be arrested and self-imaging thwarted. Alternatively, spatio-temporal spectral discretization may disrupt the propagation invariance of periodic ST wave packets and lead to self-imaging at the Talbot planes.

Here we show that aspects of both contradictory scenarios are realized by ST wave packets when endowed with a periodic spatial profile, leading to a remarkable phenomenon: a \textit{veiled} ST Talbot effect. We find that no spatial axial dynamics is discernible in the time-averaged intensity, and no axial temporal dynamics is discernible in the space-averaged intensity; Talbot revivals are altogether absent. This points to the impact of the unique spectral structure of ST wave packets trumping that of spectral discretization, whereby the initial periodic spatial profile propagates self-similarly, and may in fact display no spatial features whatsoever. Nevertheless, time-resolved measurements of the spatial profile unveil spectral-discretization-induced axial dynamics that disrupts propagation invariance: the `temporal' dynamics at every axial plane recapitulates a different portion of the `spatial' dynamics of the traditional Talbot effect, which can be shown to be a new manifestation of `time-diffraction' \cite{Longhi04OE,Porras17OL,Kondakci18PRL}. ST revivals occur at the Talbot planes with complex axial dynamics enfolding in intervening planes, which is veiled from view in physical space. Moreover, we encounter an unanticipated effect: the disparity of the transverse period observed in space from that observed in space-time. Although the spatial spectrum is sampled at multiples of $\tfrac{2\pi}{L}$, the observed period in space is nevertheless $\tfrac{L}{2}$, rather than the expected value $L$ that is observed in space-time. These phenomena associated with the veiled ST Talbot effect indicate more generally the rich repertoire of behavior intrinsic to fields endowed with precise spatio-temporal structures.

\begin{figure*}[t!]
  \begin{center}
  \includegraphics[width=14.6cm]{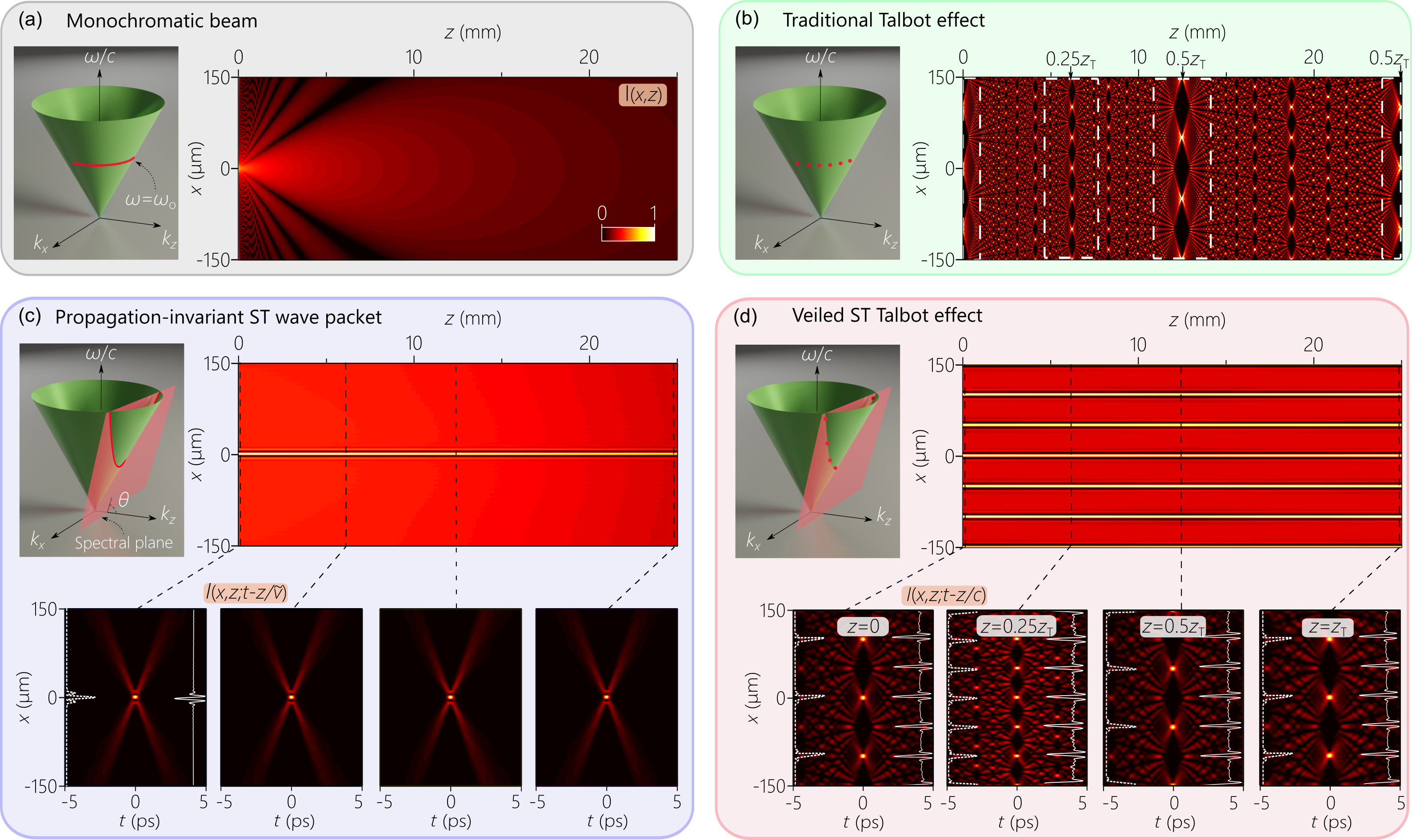}
  \end{center}
  \caption{Concept of the veiled ST Talbot effect. (a) The axial evolution of the intensity $I(x,z)$ for a monochromatic ($\lambda_{\mathrm{o}}\!=\!800$~nm) plane wave illuminating an aperture of width $\Delta x\!=\!10$~$\mu$m. On the left we depict the light-cone $k_{x}^{2}+k_{z}^{2}\!=\!(\tfrac{\omega}{c})^{2}$ and the representation of the spectral support domain of the field on its surface. (b) Plot of $I(x,z)$ for a monochromatic field $\lambda_{\mathrm{o}}\!=\!800$~nm with periodic transverse spatial profile of period $L\!=\!100$~$\mu$m and $\Delta x\!=\!10$~$\mu$m, exhibiting the traditional Talbot effect with $z_{\mathrm{T}}\!=\!25$~mm. (c) The intensity $I(x,z)$ for a ST wave packet with $\lambda_{\mathrm{o}}\!=\!800$~nm, $\Delta\lambda\!=\!2$~nm, $\Delta x\!=\!10$~$\mu$m, and $\theta\!=\!80^{\circ}$. The lower panels are the invariant spatio-temporal profiles $I(x,z;t)$ in a frame moving at $\widetilde{v}$ at selected axial planes. (d) Same as (c) except that the spatial spectrum is sampled at integer multiples of $\tfrac{2\pi}{L}$, with $L\!=\!100$~$\mu$m; $I(x,z)$ has a transverse period of $\tfrac{L}{2}\!=\!50$~$\mu$m rather than 100~$\mu$m, and there are no discernible axial changes. The lower panels are $I(x,z;t)$ in a frame moving at $c$ at selected axial planes, thereby revealing veiled periodic ST Talbot revivals. In each panel, the white dashed curve on the left is the spatial profile at $t\!=\!0$, $I(x,z;\tfrac{z}{c})$, which evolves along $z$ and is periodic with period $L$ at $z\!=\!0$, whereas the white curve on the right is the time-averaged intensity $I(x,z)$, which is independent of $z$ and is periodic with period $\tfrac{L}{2}$.}
  \label{Fig:Concept}
\end{figure*}

We first outline the theoretical basis for the veiled ST Talbot effect (see Supplementary Material). Throughout we consider for simplicity optical fields propagating along $z$ that are uniform over the transverse coordinate $y$ (i.e., a light-sheet modulated along the transverse coordinate $x$). A monochromatic paraxial field at a fixed temporal (angular) frequency $\omega\!=\!\omega_{\mathrm{o}}$ and wave number $k_{\mathrm{o}}\!=\!\tfrac{\omega_{\mathrm{o}}}{c}$ diffracts along $z$ [Fig.~\ref{Fig:Concept}(a)]. Writing the field as $E(x,z;t)\!=\!e^{i(k_{\mathrm{o}}z-\omega_{\mathrm{o}}t)}\psi_{x}(x,z)$, the envelope is given by $\psi_{x}(x,z)\!=\!\int\!dk_{x}\widetilde{\psi}(k_{x})\exp{\{i(k_{x}x-\tfrac{k_{x}^{2}}{2k_{\mathrm{o}}}z})\}$, where $k_{x}$ is the transverse component of the wave vector (denoted the spatial frequency), and the spatial spectrum $\widetilde{\psi}(k_{x})$ is the Fourier transform of $\psi_{x}(x,0)$. If the transverse field is periodic with period $L$, the spatial spectrum is discretized $k_{x}\rightarrow nk_{L}$, $\widetilde{\psi}(k_{x})\rightarrow\widetilde{\psi}(nk_{L})\!=\!\widetilde{\psi}_{n}$, and $\psi_{x}(x,z)\!=\!\sum_{n}\widetilde{\psi}_{n}\exp{\{i2\pi n(\tfrac{x}{L}-n\tfrac{z}{z_{\mathrm{T}}})\}}$; $n$ is an integer and $k_{L}\!=\!\tfrac{2\pi}{L}$. Consequently, the initial profile is revived periodically at the Talbot planes $\psi_{x}(x,mz_{\mathrm{T}})\!=\!\psi_{x}(x,0)$ for integer $m$, with rich and striking dynamics enfolding between these planes [Fig.~\ref{Fig:Concept}(b)]. Similar dynamics is observed when \textit{pulsed} fields are employed in which the spatial and temporal degrees-of-freedom are separable -- as in most mode-locked lasers, which allows $k_{x}$ to be discretized while maintaining $\Omega\!=\!\omega-\omega_{\mathrm{o}}$ continuous.

However, other pulsed-beam configurations have been recently explored in which the spatial and temporal degrees of freedom are non-separable, leading to novel emergent behaviors including propagation invariance \cite{Saari97PRL,Kondakci17NP}, controllable group velocities \cite{SaintMarie17Optica,Wong17ACSP2,Froula18NP,Kondakci19NC,Bhaduri19Optica,Jolly20OE,Bhaduri19unpublished}, transverse orbital angular momentum \cite{Chong20NP}, ultrafast beam steering \cite{Shaltout19Science}, and omni-resonant interactions with planar cavities \cite{Shiri20OL}. We consider here ST wave packets satisfying the spectral condition: $\Omega\!=\!(k_{z}-k_{\mathrm{o}})c\tan{\theta}$, which represents a plane in $(k_{x},k_{z},\tfrac{\omega}{c})$-space that makes an angle $\theta$ (the spectral tilt angle) with respect to the $k_{z}$-axis \cite{Yessenov19PRA}. the constraint $k_{x}^{2}+k_{z}^{2}\!=\!(\tfrac{\omega}{c})^{2}$ imposes a relationship between the spatial and temporal frequencies in the paraxial regime $\tfrac{\Omega}{c}(1-\cot{\theta})\!=\!\tfrac{k_{x}^{2}}{2k_{\mathrm{o}}}$, leading to a propagation-invariant (diffraction-free and dispersion-free) envelope,
\begin{equation}\label{Eq:ContinuousST}
\psi(x,z;t)=\int\!dk_{x}\widetilde{\psi}(k_{x})e^{ik_{x}x}e^{-i\Omega(t-z/\widetilde{v})}=\psi(x,0;t-z/\widetilde{v}),
\end{equation}
representing a wave packet traveling rigidly at a group velocity $\widetilde{v}\!=\!c\tan{\theta}$ [Fig.~\ref{Fig:Concept}(c)]; here $\widetilde{\psi}(k_{x})$ is the Fourier transform of $\psi(x,0,0)$. The group velocity can be tuned by tailoring the spatio-temporal spectral support domain to change $\theta$ (thus resembling aspects of tilted pulse fronts \cite{Fulop10Review}). The time-averaged intensity $I(x,z)\!=\!\int\!dt|\psi(x,z;t)|^{2}\!=\!\int\!dk_{x}|\widetilde{\psi}(k_{x})|^{2}+\int\!dk_{x}\widetilde{\psi}(k_{x})\widetilde{\psi}^{*}(-k_{x})e^{i2k_{x}x}$ as recorded by a slow detector (e.g., a camera) is independent of $z$ [Fig.~\ref{Fig:Concept}(c)]. The factor of 2 in the exponent of the second term indicates that $I(x,z)$ is scaled spatially by a factor of 2 with respect to the time-resolved intensity profile $I(x,z;t)\!=\!|\psi(x,z;t)|^{2}$, a point we will return to shortly. Finally, finite experimental resources impose an unavoidable uncertainty in the association between spatial and temporal frequencies, resulting in a `pilot' envelope accompanying the ST wave packet that travels at a group velocity of $c$ \cite{Yessenov19OE}. This pilot envelope limits the temporal window over which the ST wave packet can be observed.

When the transverse spatial profile is periodic and the spatial spectrum discretized $k_{x}\rightarrow nk_{L}$, the tight association between spatial and temporal frequencies entails that the temporal frequencies are \textit{also} discretized $\Omega\rightarrow n^{2}\Omega_{L}$, $\Omega_{L}\!=\!\tfrac{2\pi}{z_{\mathrm{T}}}\,\tfrac{c}{1-\cot{\theta}}$, and the envelope becomes
\begin{equation}\label{Eq:DiscreteST}
\psi(x,z;t)=\!\!\!\sum_{n=-\infty}^{\infty}\!\!\widetilde{\psi}_{n}e^{i2\pi n\tfrac{x}{L}}e^{-i2\pi n^{2}\tfrac{z}{z_{\mathrm{T}}}}e^{i2\pi n^{2}\tfrac{z-ct}{z_{\mathrm{T}}(1-\cot{\theta})}},
\end{equation}
where we have recast the phase factors to separate out the $(z-ct)$-term. The pilot envelope described above (accompanying the ST wave packet but traveling at $c$) is contracted after spectral discretization, so that we observe a narrower temporal window (traveling at $c$) of the ST wave packet (traveling at $\widetilde{v}$). The pilot envelope thus acts as a limited-width `searchlight' scanning the ST wave packet axially. Indeed $\psi(x,mz_{\mathrm{T}};t-m\tfrac{z_{\mathrm{T}}}{c})\!=\!\psi(x,0;t)$; that is, at the Talbot planes in a reference frame propagating at $\widetilde{v}\!=\!c$, the initial field distribution at $z\!=\!0$ is retrieved. Furthermore, comparing the formulas for the monochromatic field and the ST wave packet, we find that $\psi(x,z;t-\tfrac{z}{c})\!=\!\psi_{x}(x,z+\tfrac{ct}{1-\cot{\theta}})$. In other words, the time-resolved spatial distribution of a ST wave packet at a fixed plane $z$ corresponds to the axial spatial distribution of a \textit{monochromatic} periodic field after replacing axial displacement with a scaled time variable; which is a manifestation of `time diffraction' \cite{Longhi04OE,Porras17OL,Kondakci18PRL}; see Fig.~\ref{Fig:Concept}(d). Recording instead the time-averaged intensity $I(x,z)$ with a camera, for instance, yields
\begin{equation}\label{Eq:IntensityDiscreteST}
I(x,z)=\sum_{n}|\widetilde{\psi}_{n}|^{2}+\sum_{n}\,\,\widetilde{\psi}_{n}\widetilde{\psi}_{-n}^{*}\,\,\,\,e^{i2\cdot2\pi\tfrac{x}{L}},
\end{equation}
which indicates the complete absence of axial dynamics. Furthermore, the spatial frequency has doubled (corresponding to half the transverse spacing), so we expect a period of $\tfrac{L}{2}$ in $I(x,z)$ in contrast to a period $L$ in $I(x,z;t)$, in addition to a scaling in the size of the features in each period [Fig.~\ref{Fig:Concept}(d)].

We have carried out experiments to confirm this predicted veiled ST Talbot effect by utilizing the two-dimensional pulse-shaper developed in Refs.~\cite{Kondakci17NP,Kondakci19NC,Bhaduri19Optica} to synthesize superluminal ST wave packets at a spectral tilt angle of $\theta\!=\!80^{\circ}$, corresponding to a group velocity $\widetilde{v}\!=\!\tan{80^{\circ}}\!\approx\!5.67c$, starting with femtosecond pulses from a Ti:sapphire laser (see Supplementary Materials for details). Using a spatial light modulator, we sample the spatial spectrum at $nk_{L}$ and concurrently sample the temporal spectrum at $n^{2}\Omega_{L}$ over a temporal bandwidth of $\Delta\lambda\!\approx\!2$~nm, corresponding to temporal features of width $\Delta\tau\sim\!2.5$~ps and spatial features of width $\Delta x\sim25$~$\mu$m in the spatio-temporal profile. We measure $I(x,z)$ by scanning a CCD camera along $z$, the spatio-temporal spectrum, and $I(x,z;\tau-\tfrac{z}{c})$ measured interferometrically at different axial planes in a reference frame moving at a group velocity of $c$.

\begin{figure}[t!]
  \begin{center}
  \includegraphics[width=8.6cm]{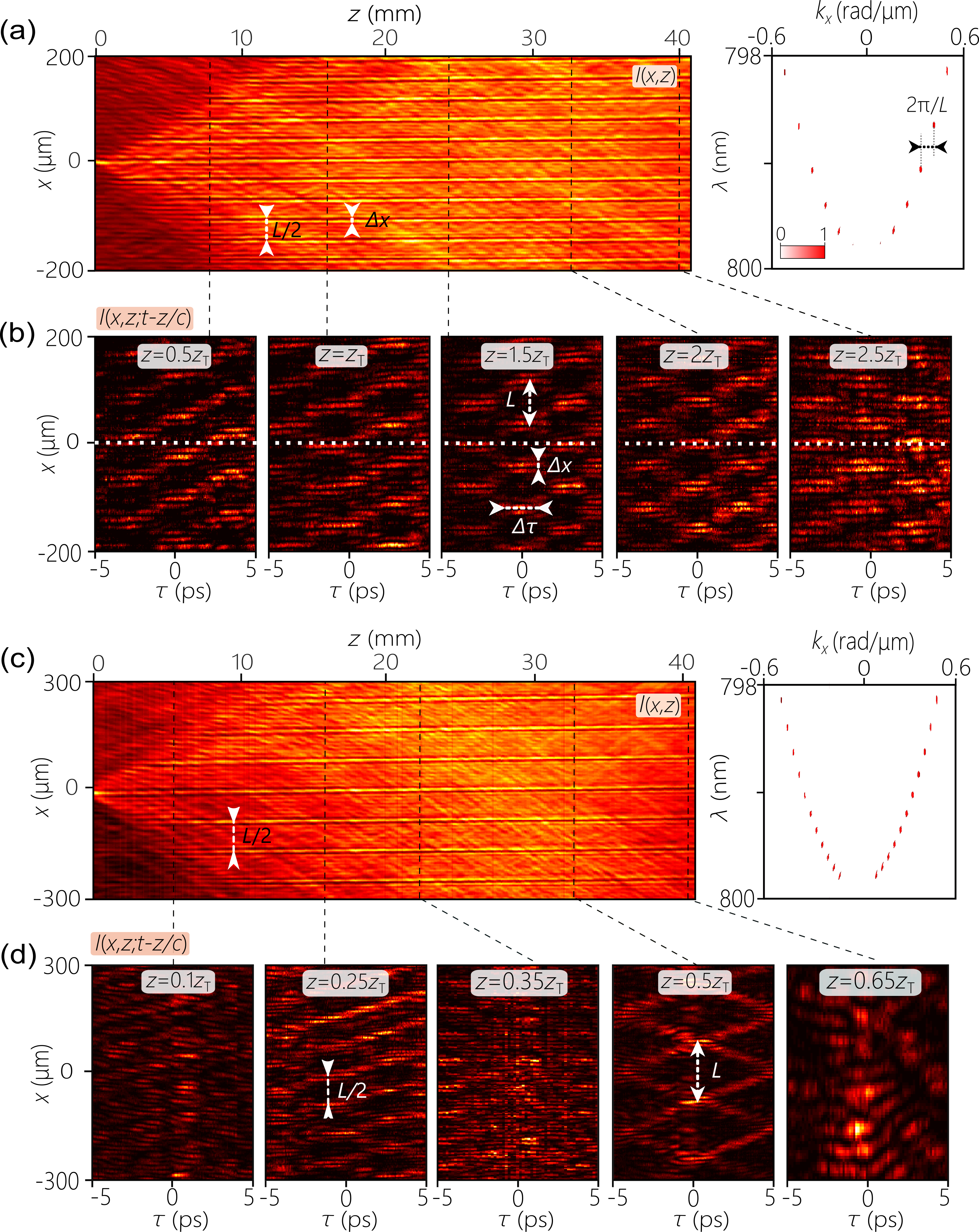}
  \end{center}
  \caption{Measurements of veiled ST Talbot revivals. (a) The measured $I(x,z)$ for a ST wave packet whose spatial spectrum is sampled at multiples of $\tfrac{2\pi}{L}$ (shown on the right); $L\!=\!80$~$\mu$m, $z_{\mathrm{T}}\!=\!16$~mm, $\theta\!=\!80^{\circ}$, $\lambda_{\mathrm{o}}\!\approx\!799.1$~nm, $\Delta\lambda\!=\!2$~nm, and $\Delta x\!=\!10$~$\mu$m. The transverse period is $\tfrac{L}{2}\!=\!40$~$\mu$m, and no axial dynamics is discernible. (b) Measured $I(x,z;\tau)$ at $z\!=\!0.5z_{\mathrm{T}}$, $z_{\mathrm{T}}$, $1.5z_{\mathrm{T}}$, $2z_{\mathrm{T}}$, and $2.5z_{\mathrm{T}}$. The transverse period is $L\!=\!80$~$\mu$m, $\Delta x\!\approx\!20-30$~$\mu$m, and $\Delta\tau\!\approx\!2.5$~ps. The white dotted line is a guide for the eye at $x\!=\!0$. (c) Same as (a) but with $L\!=\!160$~$\mu$m ($z_{\mathrm{T}}\!=\!64$~$\mu$m). The observed transverse period is $\tfrac{L}{2}\!=\!80$~$\mu$m. (d) Measured $I(x,z;\tau)$ at $z\!=\!0.1z_{\mathrm{T}}$, $0.25z_{\mathrm{T}}$, $0.35z_{\mathrm{T}}$, $0.5z_{\mathrm{T}}$, and $0.65z_{\mathrm{T}}$; $\Delta x\!\approx\!25-30$~$\mu$m, $\Delta\tau\!\approx\!2.5-3$~ps, and the measured transverse period is $L\!=\!160$~mm at $z\!=\!0.5z_{\mathrm{T}}$.}
  \label{Fig:Results1}
\end{figure}

We first sample the spatial spectrum at multiples of $\tfrac{2\pi}{L}$ with $L\!=\!80$~$\mu$m [Fig.~\ref{Fig:Results1}(a)]. Nevertheless, the measured intensity [Fig.~\ref{Fig:Results1}(a)] shows a periodic transverse profile of period $\tfrac{L}{2}\!=\!40$~$\mu$m rather than $L\!=\!80$~$\mu$m, and no trace of the Talbot effect is observable. Indeed, the axial propagation is invariant just as in previous studies of ST wave packets having continuous spectra. However, time-resolved measurements reveal a different picture altogether. The dynamics in the measured spatio-temporal profiles $I(x,z;\tau-\tfrac{z}{c})$ [Fig.~\ref{Fig:Results1}(b)] is completely veiled in the time-averaged measurements [Fig.~\ref{Fig:Results1}(a)]. We note the revivals at $z\!=\!z_{\mathrm{T}}$ and $2z_{\mathrm{T}}$, and those at $z\!=\!0.5z_{\mathrm{T}}$, $1.5z_{\mathrm{T}}$, and $2.5z_{\mathrm{T}}$, all with a transverse period of $L\!=\!80$~$\mu$m. The patterns are shifted along $x$ by $\tfrac{L}{2}$ midway between the Talbot planes, whereupon the `dark' and `bright' features are exchanged with respect to the Talbot planes. In contrast, if the spatio-temporal spectrum remains continuous without sampling, both the intensity $I(x,z)$ and the spatio-temporal profile $I(x,z;\tau)$ lack periodicity and are invariant at \textit{all} axial planes (Supplementary Material).

To resolve the dynamics within a single Talbot length, we increase the period to $L\!=\!160$~$\mu$m ($z_{\mathrm{T}}\!=\!64$~mm) by increasing the spectral sampling rate. The measured intensity $I(x,z)$ [Fig.~\ref{Fig:Results1}(c)] remains invariant with $z$, shows no axial dynamics, and has a transverse period $\tfrac{L}{2}\!=\!80$~$\mu$m rather than $L\!=\!160$~$\mu$m. The spatio-temporal profiles of the ST wave packet measured at 5 planes within a Talbot length at $z\!=\!0.15z_{\mathrm{T}}$, $0.25z_{\mathrm{T}}$, $0.35z_{\mathrm{T}}$, $0.5z_{\mathrm{T}}$, and $0.65z_{\mathrm{T}}$ reveal complex dynamics in which energy is exchanged between the peaks in the profile [Fig.~\ref{Fig:Results1}(d)]. At $z\!=\!0.5z_{\mathrm{T}}$ the transverse period is $L\!=\!160$~$\mu$m, and at $z\!=\!0.25z_{\mathrm{T}}$ is $\tfrac{L}{2}\!=\!80$~$\mu$m albeit with less contrast with respect to the Talbot planes. At intermediate planes the spatio-temporal profile takes on quasi-random distributions before returning to more recognizable profiles at Talbot half- and quarter-lengths. Once again, this dynamics is fully hidden from view in the spatial intensity [Fig.~\ref{Fig:Results1}(c)]. These experimental observations are all in excellent agreement with simulations of ST wave packets after discretizing the spatio-temporal spectrum (Supplementary Material, Supplementary Movies~1 and 2).

\begin{figure}[t]
  \begin{center}
  \includegraphics[width=8.6cm]{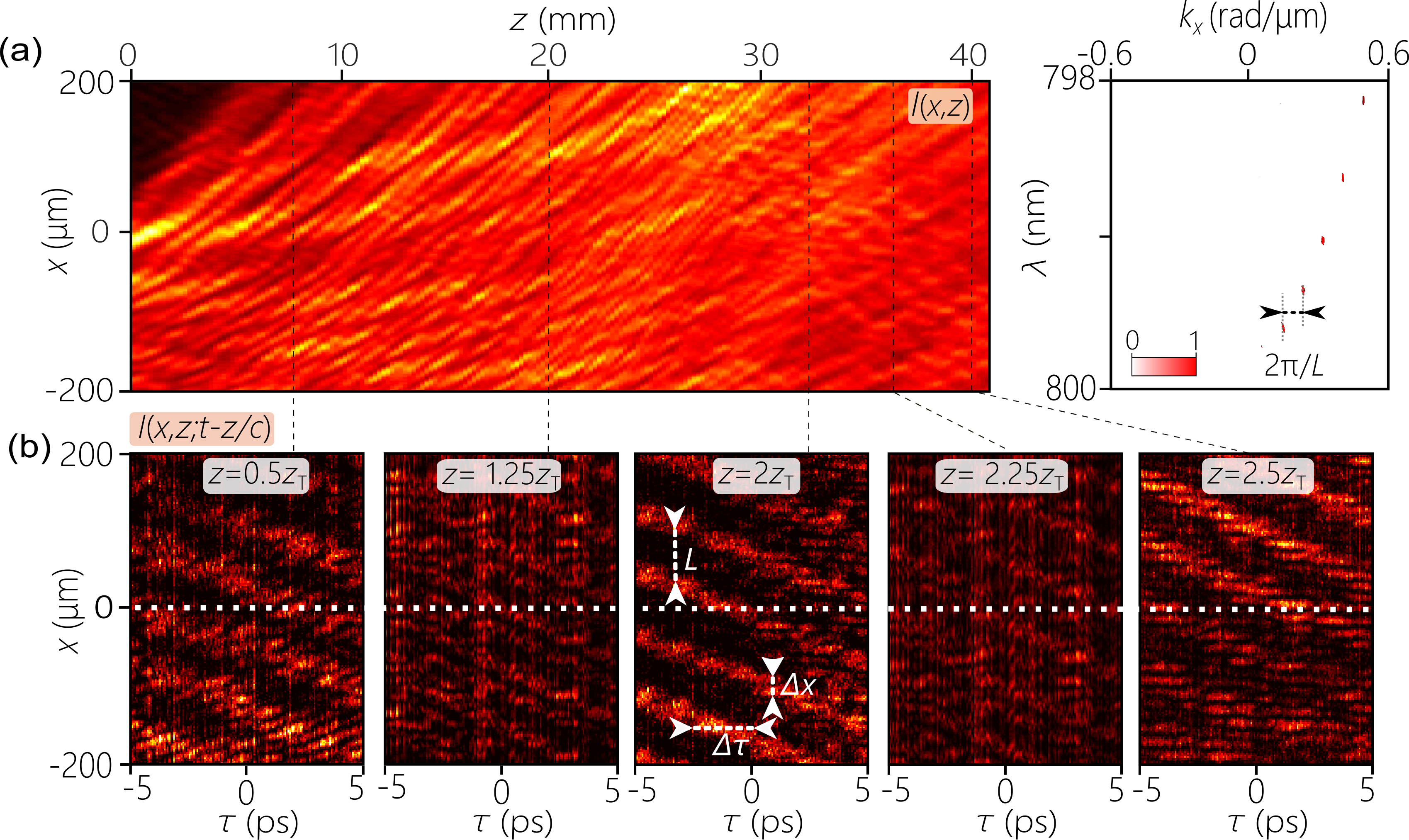}
  \end{center}
  \caption{Measurements of the veiled ST Talbot revivals for a one-sided spatial spectrum. (a) The measured intensity $I(x,z)$ with the same parameters in Fig.~\ref{Fig:Results1}(a) ($L\!=\!80$~$\mu$m) except that only positive spatial frequencies are retained (shown on right). There are no discernible transverse spatial features altogether. (b) Measured spatio-temporal profiles at $z\!=\!0.5z_{\mathrm{T}}$, $1.25z_{\mathrm{T}}$, $2z_{\mathrm{T}}$, $2.25z_{\mathrm{T}}$, and $2.5z_{\mathrm{T}}$, where $z_{\mathrm{T}}\!=\!16$~mm, $\Delta x\!\approx\!20-30$~$\mu$m, and $\Delta\tau\!\approx\!2-3$~ps.}
  \label{Fig:Results2}
\end{figure}

The veiled nature of the ST Talbot effect can be brought out even further by blocking half the spatial spectrum; $\widetilde{\psi}_{n}\!=\!0$ when $n\!<\!0$. The intensity becomes a constant $I(x,z)\!=\!\sum_{n=0}^{\infty}|\widetilde{\psi}_{n}|^{2}$ independent of $x$ and $z$. Measurements of $I(x,z)$ are shown in Fig.~\ref{Fig:Results2}(a) using the same parameters from Fig.~\ref{Fig:Results1}(a) except for the one-sided spectrum. The spatio-temporal profiles measured at different axial planes reveal the underlying periodic transverse axial structure that is altogether veiled in the intensity and verify the axial dynamics associated with time-diffraction. The ST Talbot effect is seen in the self-imaging revivals at $z\!=\!1.25z_{\mathrm{T}}$ and $2.25z_{\mathrm{T}}$, and at $z\!=\!0.5z_{\mathrm{T}}$ and $2.5z_{\mathrm{T}}$ (see Supplementary Movie~3).

The factor of 2 in the time-averaged intensity (Eq.~\ref{Eq:IntensityDiscreteST}) with respect to spatio-temporal envelope (Eq.~\ref{Eq:DiscreteST}) appears regularly in correlation functions in the context of optical coherence after averaging over a statistical ensemble. However, the underlying stochastic fields are inaccessible, and the correlation functions are the only observables. Here, we capture the intensity after time-averaging (which is entangled with the spatial degree of freedom \cite{Kondakci19OL}) \textit{and} the underlying spatio-temporal profile prior to time-averaging, thus allowing an unambiguous observation of the impact of this spatial-scaling factor in changing the transverse period in the two distinct domains. 

The unique behaviors observed and recent advances achieved using ST wave packets rely on inculcating a precise spatio-temporal structure into the optical field. The current work raises the question regarding which features of ST wave packets survive spectral discretization. We have shown that propagation invariance is maintained in the time-averaged intensity after spectral discretization. Nevertheless, the underlying field exhibits complex dynamics in space-time associated with time-diffraction that reveals self-imaging Talbot revivals, all of which is completely veiled in the spatial intensity profile. The time-resolved dynamics reported here is reminiscent of quantum carpets \cite{Kaplan98PS} despite the absence of an index-confining structure. Although we have couched this phenomenon in terms of optical fields, it is applicable to any other waves, such as acoustics or matter waves. 

\noindent
\textbf{Acknowledgments.} This work was funded by the U.S. Office of Naval Research, contract N00014-17-1-2458.

\bibliography{diffraction}

\end{document}

\preprint{APS/123-QED}

\title{Veiled Talbot effect: Supplementary material}

\author{Murat Yessenov$^{1}$}
\thanks{These authors contributed equally to this work}
\author{Layton A. Hall$^{1}$}
\thanks{These authors contributed equally to this work}
\author{Sergey A. Ponomarenko$^{2,3}$}
\author{Ayman F. Abouraddy$^{1}$}
%\email{raddy@creol.ucf.edu}
\affiliation{$^{1}$CREOL, The College of Optics \& Photonics, University of Central~Florida, Orlando, FL 32816, USA\\$^{2}$Department of Elect. and Computer Eng., Dalhousie University, Halifax, Nova Scotia B3J 2X4, Canada\\$^{3}$Department of Physics and Atmospheric Science, Dalhousie University, Halifax, Nova Scotia B3H 4R2, Canada}

%\begin{abstract}
%bbbb
%\end{abstract}

\renewcommand{\thesection}{S\arabic{section}}   
\renewcommand{\thefigure}{Supplementary Figure \arabic{figure}}
\renewcommand{\theequation}{S\arabic{equation}}  

\maketitle

%\tableofcontents

%\newpage

\section{Traditional Talbot effect}

\subsection{Monochromatic field}

Consider a monochromatic field at a temporal frequency $\omega\!=\!\omega_{\mathrm{o}}$,  $E(x,z;t)\!=\!e^{i(k_{\mathrm{o}}z-\omega_{\mathrm{o}}t)}\psi_{x}(x,z)$, with an envelope given by $\psi_{x}(x,z)=\int\!dk_{x}\widetilde{\psi}(k_{x})e^{ik_{x}x}e^{i(k_{z}-k_{\mathrm{o}})z}$; here the spatial spectrum $\widetilde{\psi}(k_{x})$ is the Fourier transform of $\psi_{x}(x,0)$, $k_{\mathrm{o}}\!=\!\tfrac{\omega_{\mathrm{o}}}{c}\!=\!\tfrac{2\pi}{\lambda_{\mathrm{o}}}$ is the wave number corresponding to the temporal frequency $\omega_{\mathrm{o}}$ (wavelength $\lambda_{\mathrm{o}}$), and $c$ is the speed of light in vacuum. The dispersion relationship in free space enforces the relationship $k_{z}^{2}\!=\!k_{\mathrm{o}}^{2}-k_{x}^{2}$. In the paraxial limit, we have $k_{z}-k_{\mathrm{o}}\!\approx\!-\tfrac{k_{x}^{2}}{2k_{\mathrm{o}}}$, and the envelope takes the form:
\begin{equation}\label{Eq:ContinuousDiffraction}
\psi_{x}(x,z)=\int\!dk_{x}\widetilde{\psi}(k_{x})e^{ik_{x}x}e^{-i\tfrac{k_{x}^{2}}{2k_{\mathrm{o}}}z}.
\end{equation}

If the field at $z\!=\!0$ is periodic along $x$ with period $L$, $\psi_{x}(x+mL,0)\!=\!\psi_{x}(x,0)$ for integer $m$, then the spatial spectrum is discrete and sampled at periods $k_{L}\!=\!\tfrac{2\pi}{L}$, so that $k_{x}\!\rightarrow\!nk_{L}$ and $\widetilde{\psi}(k_{x})\!\rightarrow\!\widetilde{\psi}(nk_{L})\!=\!\widetilde{\psi}_{n}$, where $n$ is an integer, and the envelope becomes
\begin{equation}\label{Eq:DiscreteDiffraction}
\psi_{x}(x,z)=\sum_{n=-\infty}^{\infty}\widetilde{\psi}_{n}e^{i2\pi n\tfrac{x}{L}}e^{-i2\pi n^{2}\tfrac{z}{z_{\mathrm{T}}}},
\end{equation}
where $z_{\mathrm{T}}\!=\!\tfrac{2L^{2}}{\lambda_{\mathrm{o}}}$ is the Talbot length. Therefore, whenever $z\!=\!mz_{\mathrm{T}}$ (for integer $m$), we have
\begin{equation}
\psi_{x}(x,mz_{\mathrm{T}})=\sum_{n=-\infty}^{\infty}\widetilde{\psi}_{n}e^{i2\pi n\tfrac{x}{L}}=\psi_{x}(x,0).
\end{equation}
That is, at the Talbot planes $z\!=\!mz_{\mathrm{T}}$, periodic revivals of the initial field distribution occur. The intensity $I(x,z)\!=\!|\psi_{x}(x,z)|^{2}$ has the same period $L$ along $x$ at all the Talbot planes. The evolution of the Talbot revivals are explicit in physical space.

\subsection{Pulsed field}

If the field is pulsed $E(x,z;t)\!=\!e^{i(k_{\mathrm{o}}z-\omega_{\mathrm{o}}t)}\psi(x,z;t)$, then the envelope takes the form
\begin{equation}
\psi(x,z;t)=\iint\!dk_{x}d\Omega\:\:\:\widetilde{\psi}(k_{x},\Omega)e^{ik_{x}x}e^{i(k_{z}-k_{\mathrm{o}})z}e^{-i\Omega t},
\end{equation}
where $\Omega\!=\!\omega-\omega_{\mathrm{o}}$ is the temporal frequency measured respect to the fixed frequency $\omega_{\mathrm{o}}$, and the spatio-temporal spectrum $\widetilde{\psi}(k_{x},\Omega)$ is the 2D Fourier transform of $\psi(x,0;t)$ with respect to $x$ and $t$. For most laser pulses we can make the assumption that the spatio-temporal spectrum is separable with respect to the spatial and temporal degrees of freedom $\widetilde{\psi}(k_{x},\Omega)\!\approx\!\widetilde{\psi}_{x}(k_{x})\widetilde{\psi}_{t}(\Omega)$, in which case $\psi(x,0;t)\!\approx\!\psi_{x}(x)\psi_{t}(t)$, where $\widetilde{\psi}_{x}(k_{x})$ if the Fourier transform of $\psi_{x}(x)$ and $\widetilde{\psi}_{t}(\Omega)$ is the Fourier transform of $\psi_{t}(t)$. Besides the separability of the field in space and time at $z\!=\!0$, the field remains approximately separable along $z$ if space-time coupling remains minimal, which requires narrow spatial and temporal bandwidths. Within these limits, we have $k_{z}-k_{\mathrm{o}}\!\approx\!\tfrac{\Omega}{c}-\tfrac{k_{x}^{2}}{2k_{\mathrm{o}}}$, leading to an envelope of the form
\begin{eqnarray}\label{Eq:TraditionalPulsedBeam}
\psi(x,z;t)\!&=&\!\!\int\!\!dk_{x}\widetilde{\psi}(k_{x})e^{ik_{x}x}e^{-i\tfrac{k_{x}^{2}}{2k_{\mathrm{o}}}z}\!\!\int\!\!d\Omega\widetilde{\psi}_{t}(\Omega)e^{-i\Omega(t-z/c)}\nonumber\\&=&\psi_{x}(x,z)\psi_{t}(t-z/c),
\end{eqnarray}
where the term $\psi_{x}(x,z)$ is the same as that for a monochromatic beam in Eq.~\ref{Eq:ContinuousDiffraction}, and $\psi_{t}(t-z/c)$ is a pulse temporal envelope traveling at a group velocity $c$.

Because of the separability of the spatio-temporal spectrum with respect to $k_{x}$ and $\Omega$, we can discretize the spatial spectrum along $k_{x}$ while maintaining $\Omega$ a continuous variable. Introducing a periodic spatial structure with period $L$ along $x$, and utilizing the same nomenclature from the previous subsection, results in a spatial envelope given by
\begin{equation}
\psi_{x}(x,z)=\sum_{n=-\infty}^{\infty}\widetilde{\psi}_{n}e^{i2\pi n\tfrac{x}{L}}e^{-i2\pi n^{2}\tfrac{z}{z_{\mathrm{T}}}},
\end{equation}
which is identical to that of the periodic monochromatic field in Eq.~\ref{Eq:DiscreteDiffraction}. Consequently, $\psi(x,mz_{\mathrm{T}};t)\!=\!\psi_{x}(x,0)\psi_{t}(t-mz_{\mathrm{T}}/c)$; that is, the transverse periodic spatial pattern undergoes revivals at the Talbot planes accompanied by a temporal envelope propagating at a group velocity $c$ independently of the spatial evolution of the field. Once again, the spatial evolution associated with the Talbot effect is explicit in physical space for pulsed fields.

\section{Space-time Talbot effect}

\subsection{ST wave packets with continuous spatio-temporal spectrum}

Because of the tight association between the spatial and temporal frequencies intrinsic to ST wave packets, the spatio-temporal spectrum is no longer a function of $k_{x}$ and $\Omega$. Rather, its dimensionality is reduced, and the envelope takes the form
\begin{equation}
\psi(x,z;t)=\int\!dk_{x}\widetilde{\psi}(k_{x})e^{ik_{x}x}e^{-i(k_{z}-k_{\mathrm{o}})z}e^{-i\Omega t},
\end{equation}
where $\Omega$ is no longer an independent variable, but depends on $k_{x}$. There are two constraints on the values of $k_{x}$, $k_{z}$ and $\Omega$: (1) the dispersion relationship in free space $k_{x}^{2}+k_{z}^{2}\!=\!(\tfrac{\omega}{c})^{2}$, and (2) the definition of the tilted spectral plane $\Omega\!=\!(k_{z}-k_{\mathrm{o}})c\tan{\theta}$, where $\theta$ is the spectral tilt angle made by the plane with respect to the $k_{z}$-axis. Therefore, the envelope can be written as
\begin{equation}\label{Eq:ContinuousSTWavePacket}
\psi(x,z;t)=\int\!dk_{x}\widetilde{\psi}(k_{x})e^{ik_{x}x}e^{-i\Omega(t-z/\widetilde{v})}=\psi(x,0;t-z/\widetilde{v}).
\end{equation}
The wave packet travels rigidly in free space at a group velocity $\widetilde{v}\!=\!c\tan{\theta}$.

In anticipation of the discretization of $k_{x}$ when considering the ST Talbot effect, we express the envelope by an integral with the integrand expressed explicitly in terms of $k_{x}$. Implementing the same paraxial and narrowband approximations as in the traditional pulsed case described above, we have $k_{z}-k_{\mathrm{o}}\!\approx\!\tfrac{\Omega}{c}-\tfrac{k_{x}^{2}}{2k_{\mathrm{o}}}$. The additional constraint associated with the spectral plane can be used to eliminate $k_{z}$, so that we have
\begin{equation}\label{Eq:OmegaKxExpression}
\frac{\Omega}{c}(1-\cot{\theta})=\frac{k_{x}^{2}}{2k_{\mathrm{o}}}.
\end{equation}
For a plane-wave pulse ($k_{x}\!=\!0$, $\theta\!=\!45^{\circ}$), the dispersion relationship is the light-line $\Omega\!=\!c(k_{z}-k_{\mathrm{o}})$, which indicates that the group velocity is $c$. The existence of a spatial spectrum ($k_{x}\!\neq\!0$, $\theta\!\neq\!45^{\circ}$) leads to deviation away from this light-line, and the tight association between the spatial frequencies and temporal frequencies according to Eq.~\ref{Eq:OmegaKxExpression} changes the group velocity from $\widetilde{v}\!=\!c$ to $\widetilde{v}\!=\!c\tan{\theta}$, as seen in Eq.~\ref{Eq:ContinuousSTWavePacket}. We now rewrite the envelope of the ST wave packet explicity in terms of $k_{x}$,
\begin{equation}
\psi(x,z;t)=\int\!dk_{x}\,\,\widetilde{\psi}(k_{x})\,\,e^{ik_{x}x}\,\,e^{-i\tfrac{k_{x}^{2}}{2k_{\mathrm{o}}}z}\,\,e^{i\tfrac{k_{x}^{2}}{2k_{\mathrm{o}}}\tfrac{1}{1-\cot{\theta}}(z-ct)}.
\end{equation}
In contrast to Eq.~\ref{Eq:TraditionalPulsedBeam} that applies to a traditional pulsed beam, the integral here does \textit{not} separate into a product of functions of space and time. In anticipation of the analysis below, note that the exponent with the term $\tfrac{k_{x}^{2}}{2k_{\mathrm{o}}}$, which is responsible for the change in the group velocity, will be neutralized at the Talbot planes once the spatial spectrum is discretized. We thus expect that the group velocity of a ST wave packet with periodic transverse distribution will no longer be $c\tan{\theta}$, but will revert instead to $c$. 

\begin{figure}[t!]
  \begin{center}
  \includegraphics[width=7.6cm]{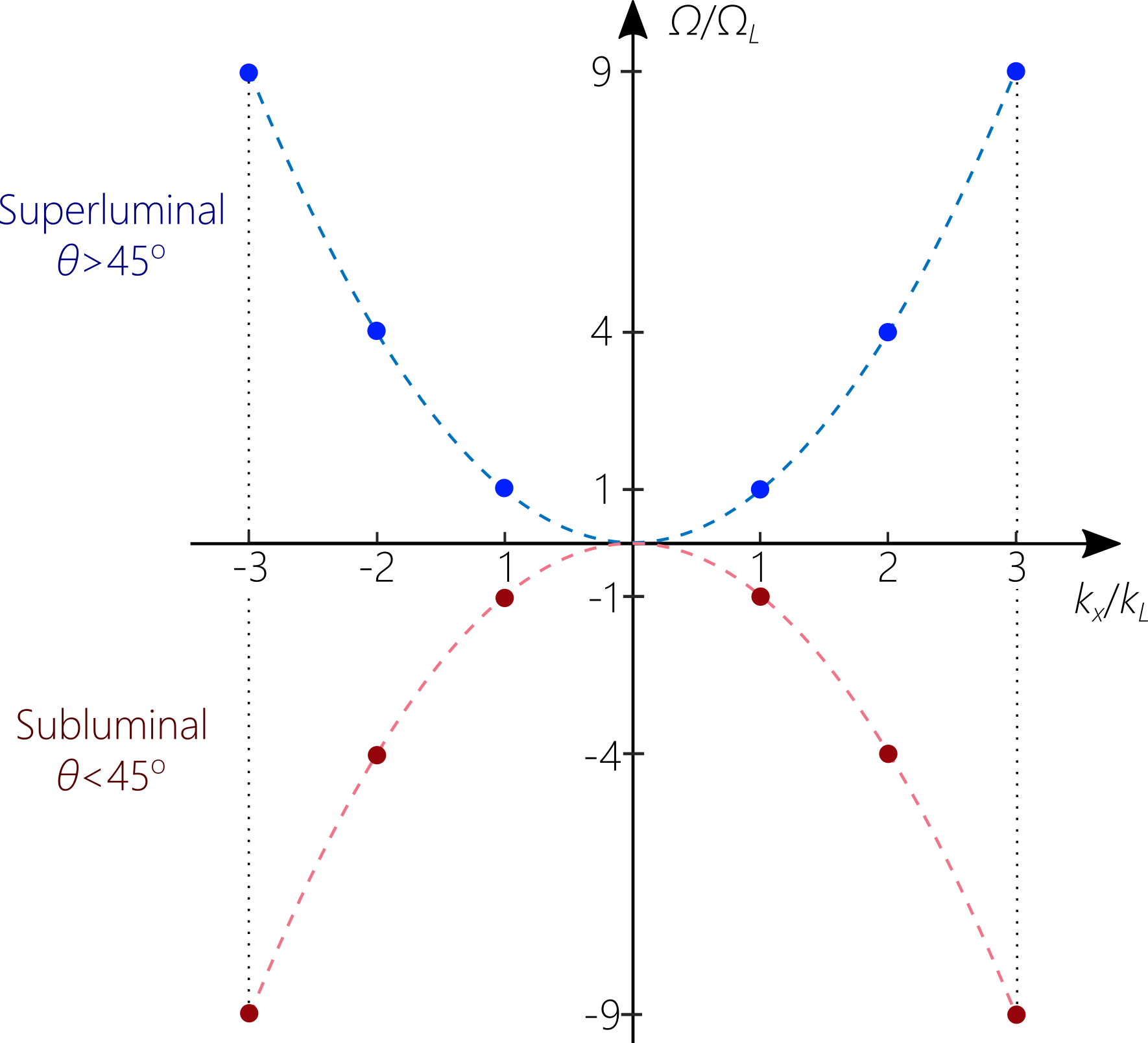}
  \end{center}
  \caption{Spatio-temporal spectral support domain of subluminal and superluminal ST wave packets projected onto the $(k_{x},\tfrac{\Omega}{c})$-plane. The spectrum is discretized $k_{x}\rightarrow nk_{L}$ and $\Omega\rightarrow n^{2}\Omega_{L}$.}
  \label{Fig:Discretization}
\end{figure}

We have two domains of interest that must be delineated here (see Fig.~\ref{Fig:Discretization}). The first is the subluminal regime $\theta\!<\!45^{\circ}$, whereupon $\widetilde{v}\!<\!c$ and the term $1-\cot{\theta}$ is negative, so that Eq.~\ref{Eq:OmegaKxExpression} enforces $\Omega\!<\!0$. In other words, $\omega_{\mathrm{o}}$ is the maximum temporal frequency in the spectrum of a subluminal ST wave packet ($\omega\!<\!\omega_{\mathrm{o}}$). The second regime is superluminal with $\theta\!>\!45^{\circ}$, $\widetilde{v}\!>\!c$, and $1-\cot{\theta}\!>\!0$, so that Eq.~\ref{Eq:OmegaKxExpression} enforces $\Omega\!>\!0$. In other words, $\omega_{\mathrm{o}}$ is the minimum temporal frequency in the spectrum of a superluminal ST wave packet ($\omega\!>\!\omega_{\mathrm{o}}$).

We have assumed perfect correlations between the spatial and temporal frequencies; that is, each spatial frequency $k_{x}$ is ideally associated with a single temporal frequency $\omega$ (or wavelength $\lambda$). Such an ideal scenario requires infinite energy \cite{Sezginer85JAP}. Any finite realistic system producing a finite-energy ST wave packet introduces an unavoidable `fuzziness' in the association between the spatial and temporal frequencies that we term the spectral uncertainty $\delta\omega$ ($\delta\lambda$ on a wavelength scale) \cite{Yessenov19OE}; the spectral uncertainty is typically much smaller than than the temporal bandwidth $\delta\lambda\!\ll\!\Delta\lambda$. As a consequence, the ST wave packet has a finite propagation distance \cite{Kondakci19OL}. The realistic finite-energy ST wave packet can be decomposed into a product of an ideal ST wave packet traveling at a group velocity $\widetilde{v}$ and a so-called `pilot' wave packet that travels at a group velocity $c$. The temporal width of this pilot envelope is the inverse of the temporal uncertainty $\delta\omega$ \cite{Yessenov19OE}. Because of the difference in group velocities of the ST wave packet and the pilot envelope,$\widetilde{v}\!=\!c\tan{\theta}$ and $c$, respectively, the propagation distance $L_{\mathrm{max}}$ of the ST wave packet is related to this difference and the spectral uncertainty \cite{Yessenov19OE}. 

Observations made by a camera (or other `slow' detector that cannot resolve the ST wave packet in time) scanned along the propagation direction $z$ correspond to the time-averaged intensity $I(x,z)\!=\!\int\!dt|\psi(x,z;t)|^{2}$, which takes the form
\begin{equation}\label{Eq:ContinuousTimeAvergaedST}
I(x,z)=\int\!dk_{x}\,|\widetilde{\psi}(k_{x})|^{2}\,+\,\int\!dk_{x}\,\widetilde{\psi}(k_{x})\widetilde{\psi}^{*}(-k_{x})e^{i2k_{x}x}.
\end{equation}
Several pertinent observations with regards to Eq.~\ref{Eq:ContinuousTimeAvergaedST} will be needed subsequently. First, Eq.~\ref{Eq:ContinuousTimeAvergaedST} is altogether independent of $z$; i.e., there are no observable axial dynamics in the time-averaged intensity evolution. The precise association between the axial wave numbers and temporal frequencies according to $\Omega\!=\!(k_{z}-k_{\mathrm{o}})c\tan{\theta}$ lead to the time-averaging process wiping out any $z$-dependence. Second, the first term in Eq.~\ref{Eq:ContinuousTimeAvergaedST} is a constant background term that is independent of the field distribution. Third, the second term in Eq.~\ref{Eq:ContinuousTimeAvergaedST} is responsible for the transverse spatial variation in the time-averaged intensity. If the spatial spectrum is an even function $\widetilde{\psi}(-x)\!=\!\widetilde{\psi}(x)$, then the field takes the form of a peaked function atop the background term; whereas an odd function $\widetilde{\psi}(-k_{x})\!=\!\widetilde{\psi}(k_{x})$ produces a dip below the background. Fourth, note the factor of 2 that appears in the exponent in the second term in Eq.~\ref{Eq:ContinuousTimeAvergaedST}. This factor signifies a reduction of transverse spatial scale by 2, which routinely appears in coherence theory. However, because the underlying field is not observable when using incoherent sources, this factor has not been of critical importance to date. We show below that this factor of 2 leads to an interesting feature of the ST Talbot effect. Finally, if the spatial spectrum is one-sided (i.e., $\widetilde{\psi}(k_{x})\!=\!0$ when $k_{x}\!<\!0$), then the time-averaged intensity is reduced simply to the constant background $I(x,z)\!=\!\int_{0}^{\infty}\!dk_{x}|\widetilde{\psi}(k_{x})|^{2}$ with no observable spatial variation. 

\subsection{ST wave packets with discretized spectrum}

If a transverse periodic spatial pattern along $x$ with period $L$ is imposed at $z\!=\!0$, then the spatial spectrum is sampled at $k_{x}\rightarrow nk_{L}$ where $k_{L}\!=\!\tfrac{2\pi}{L}$ as before. However, because the spatial and temporal frequencies are related, discretizing the spatial frequencies $k_{x}$ entails that the temporal frequencies $\Omega$ are also discretized. However, because $\Omega$ is related to $k_{x}$ quadratically as shown in Eq.~\ref{Eq:OmegaKxExpression}, it is sampled at $\Omega\rightarrow n^{2}\Omega_{L}$, where $\Omega_{L}\!=\!\tfrac{2\pi}{z_{\mathrm{T}}}\,\tfrac{c}{1-\cot{\theta}}$. The envelope of the wave packet now takes the form:
\begin{equation}
\psi(x,z;t)=\!\!\sum_{n=-\infty}^{\infty}\!\!\!\widetilde{\psi}_{n}e^{i2\pi n\tfrac{x}{L}}e^{-i2\pi n^{2}\tfrac{z}{z_{\mathrm{T}}}}e^{i2\pi n^{2}\tfrac{1}{1-\cot{\theta}}\tfrac{z-ct}{z_{\mathrm{T}}}}.
\end{equation}
Note that the term $e^{-i2\pi n^{2}\tfrac{z}{z_{\mathrm{T}}}}$ vanishes at the Talbot planes $z\!=\!mz_{\mathrm{T}}$, which indicates (as anticipated above) that the wave packet propagates at a group velocity $\widetilde{v}\!=\!c$ rather than $\widetilde{v}\!=\!c\tan{\theta}$ as for its continuous-spectrum counterpart.

We have separated out the term containing $z-ct$ in the equation above. We show below in the description of the phase patterns for synthesizing periodic ST wave packets that the spectral discretization increases the spectral uncertainty $\delta\omega$, thereby reducing the width of the pilot envelope. The observed spatio-temporal profile will thus be the portion of the ST wave packet revealed by this narrower pilot envelope \cite{Yessenov19OE}. Writing the ST wave packet envelope in terms of $z-ct$ therefore highlights the observable spatio-temporal profile.

At the Talbot planes ($z\!=\!mz_{\mathrm{T}}$), the envelope in a reference frame traveling at a group velocity $\widetilde{v}\!=\!c$ ($ct\!=\!mz_{\mathrm{T}}$) takes the form
\begin{equation}
\psi(x,mz_{\mathrm{T}};\tfrac{mz_{\mathrm{T}}}{c})=\sum_{n=-\infty}^{\infty}\widetilde{\psi}_{n}e^{i2\pi n\tfrac{x}{L}}=\psi(x,0;0),
\end{equation}
indicating perfect revivals of the initial field at the Talbot planes. The envelope observed in time at the Talbot planes ($ct\!=\!mz_{\mathrm{T}}+c\tau$) has the form
\begin{eqnarray}
\psi(x,mz_{\mathrm{T}};\tfrac{mz_{\mathrm{T}}}{c}+\tau)&=&\!\!\sum_{n=-\infty}^{\infty}\!\!\!\widetilde{\psi}_{n}e^{i2\pi n\tfrac{x}{L}}e^{-i2\pi n^{2}\tfrac{1}{1-\cot{\theta}}\tfrac{c\tau}{z_{\mathrm{T}}}}\nonumber\\&=&\psi(x,0;\tau),
\end{eqnarray}
so that the time-resolved profile at the Talbot planes (in the moving reference frame) corresponds to the profile at the initial plane $z\!=\!0$. Note that this equation is identical to the Talbot effect effect as displayed by the monochromatic field or the spatial part of the pulsed field (when the spatial and temporal degrees of freedom are separable),
\begin{equation}
\psi\left(x,z;\frac{z}{c}+\tau\right)=\psi_{x}\left(x,z+\frac{c\tau}{1-\cot{\theta}}\right).
\end{equation}
The axial coordinate $z$ is replaced by $z\!=\!\tfrac{c\tau}{1-\cot{\theta}}$. In other words, the profile observed in space along the axis in the traditional Talbot configuration is now observed in the time domain at fixed axial planes.

The time-averaged intensity $I(x,z)\!=\!\int\!dt|\psi(x,z;t)|^{2}$ is given by
\begin{equation}
I(x,z)=\sum_{n=-\infty}^{\infty}|\widetilde{\psi}_{n}|^{2}+\sum_{n-\infty}^{\infty}\widetilde{\psi}_{n}\widetilde{\psi}_{-n}^{*}\:\:e^{i2\cdot2\pi n\tfrac{x}{L}}.
\end{equation}
Note the absence of any dependence on $z$, so that the underlying complex dynamics associated with time-diffraction is altogether veiled in physical space. Also note the appearance of the factor of 2 in the exponent, which indicates that the transverse spatial period observed in physical space is $L/2$ rather than $L$. Therefore, although the spatial spectrum is sampled at integer multiples of $k_{L}\!=\!\tfrac{2\pi}{L}$ and the time-resolved measurements of the envelope profile reveals a periodic distribution of period $L$, observing the field with a slow detector reveals a periodic distribution of period $\tfrac{L}{2}$. In other words, twice the number of peaks will be observed in the time-averaged intensity than in the time-resolved profile, with half of those peaks not corresponding to underlying time-resolved peaks. Finally, when the spatial spectrum is one-sided ($\widetilde{\psi}_{n}\!=\!0$ if $n\!<\!0$), then the time-averaged intensity is a constant $I(x,z)\!=\!\sum_{n=0}^{\infty}|\widetilde{\psi}_{n}|^{2}$ with no observable spatial variations.

\begin{figure}[t!]
  \begin{center}
  \includegraphics[width=8.6cm]{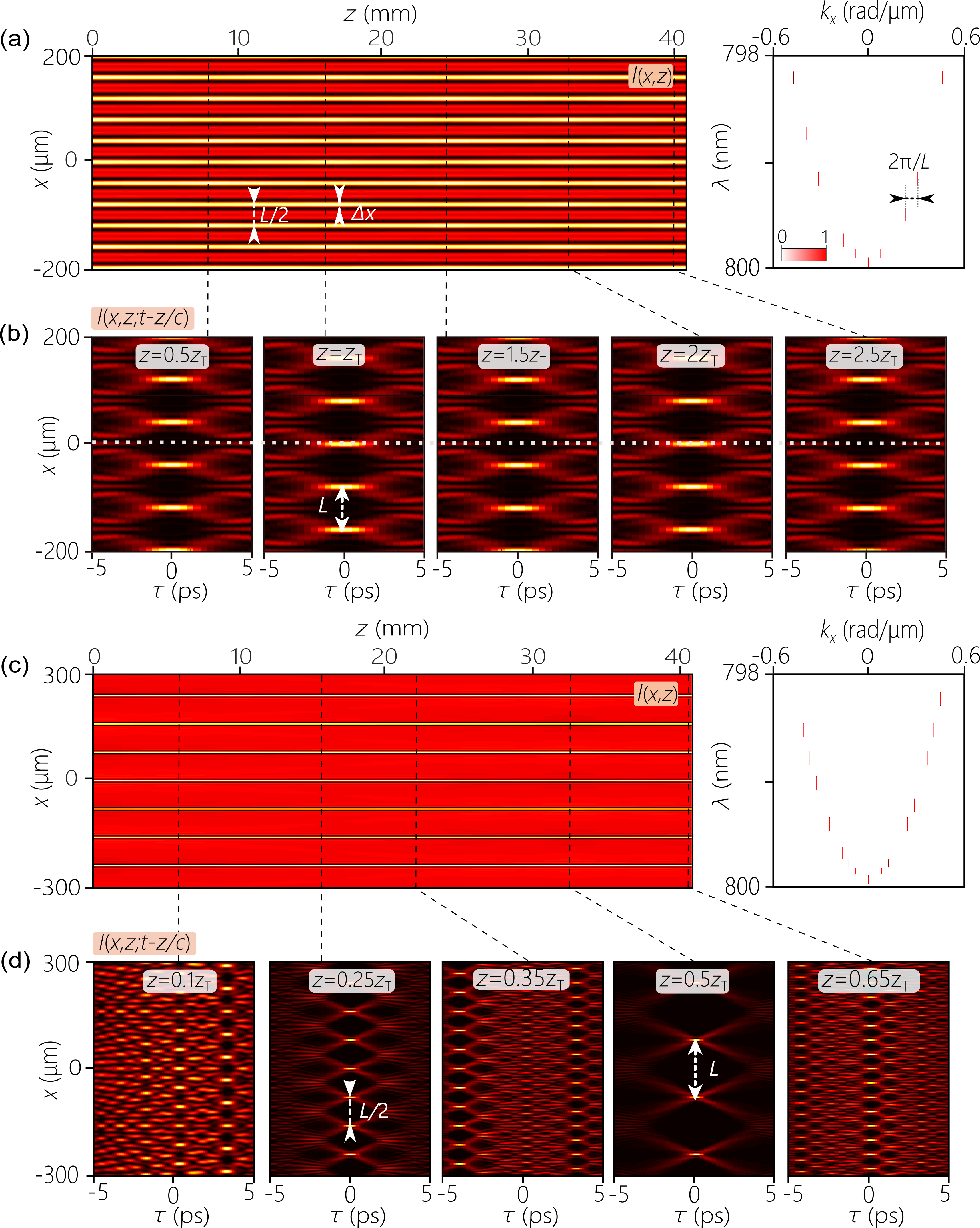}
  \end{center}
  \caption{(a) Calculated time-averaged intensity $I(x,z)$ using the same experimental parameters from Fig.~2(a) in the main text: $\Delta\lambda\!=\!2$~nm, $\theta\!=\!80^{\circ}$, $L\!=\!80$~$\mu$m, and $\Delta x\!=\!10$~$\mu$m. (b) The spatio-temporal profiles calculated at particular axial planes in (a) extending over two Talbot lengths. The dotted white line is a guide for the eye. (c) Same as (a), but using the same parameters as in Fig.~2(c) in the main text: $L\!=\!160$~$\mu$m and $\Delta x\!=\!8$~$\mu$m. (d) The spatio-temporal profiles calculated at particular axial planes in (c) within a single Talbot period.}
  \label{Fig:SimulationsPeriodic}
\end{figure}

\begin{figure}[t!]
  \begin{center}
  \includegraphics[width=8.6cm]{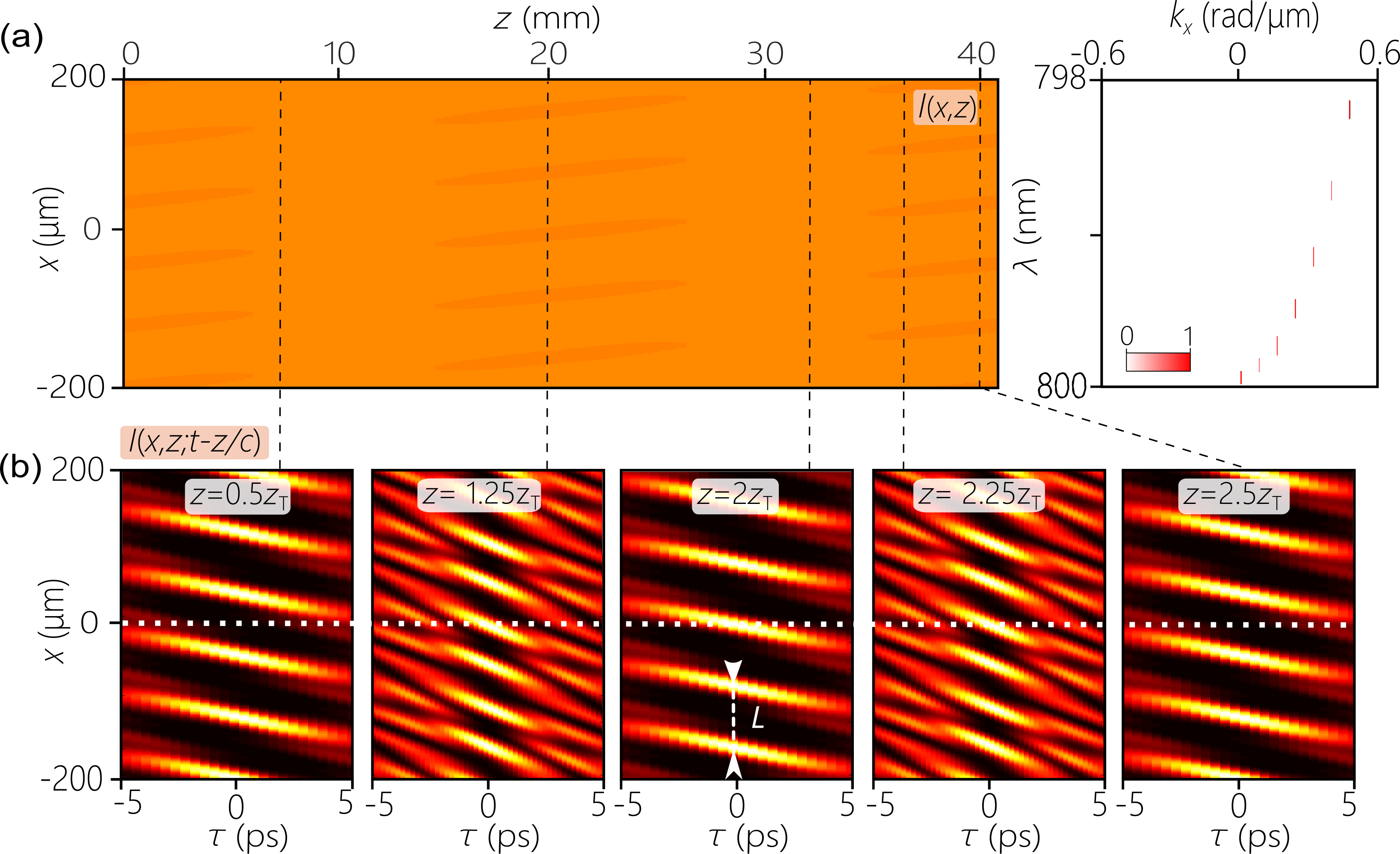}
  \end{center}
  \caption{(a) Calculated time-averaged intensity $I(x,z)$ using the same experimental parameters from Fig.~3(a) in the main text: $\Delta\lambda\!=\!2$~nm, $\theta\!=\!80^{\circ}$, $L\!=\!80$~$\mu$m, and $\Delta x\!=\!14$~$\mu$m. (b) The spatio-temporal profiles calculated at particular axial planes in (a) extending over two Talbot lengths. The dotted white line is a guide for the eye.}
  \label{Fig:SimulationsConstant}
\end{figure}

\section{Simulations of the space-time Talbot effect}

\subsection{Simulations corresponding to the configurations in the main text}

We have calculated the time-averaged intensity $I(x,z)$ and the spatio-temporal profile $I(x,z;t)$ at the axial planes corresponding to the measurements reported in Fig.~2 and Fig.~3 in the main text. We make use of the theoretical model for ST wave packets developed in Refs.~\cite{Yessenov19PRA,Yessenov19OE}, which includes the finite spectral uncertainty that sets a limit on the propagation distance. The simulations corresponding to Fig.~2 in the main text in which a periodic profile is imposed on the ST wave packet through spectral discretization are plotted in \ref{Fig:SimulationsPeriodic}. The simulations corresponding to Fig.~3 in the main text in which only one half of the discretized spatial spectrum is taken into consideration, thus eliminating and spatial features in the time-averaged intensity, are plotted in \ref{Fig:SimulationsConstant}. In all cases, we observe excellent agreement between the experimental observations and the simulation results.

\subsection{Simulations for the effect of the spectral tilt angle}

\begin{figure}[t!]
  \begin{center}
  \includegraphics[width=8.6cm]{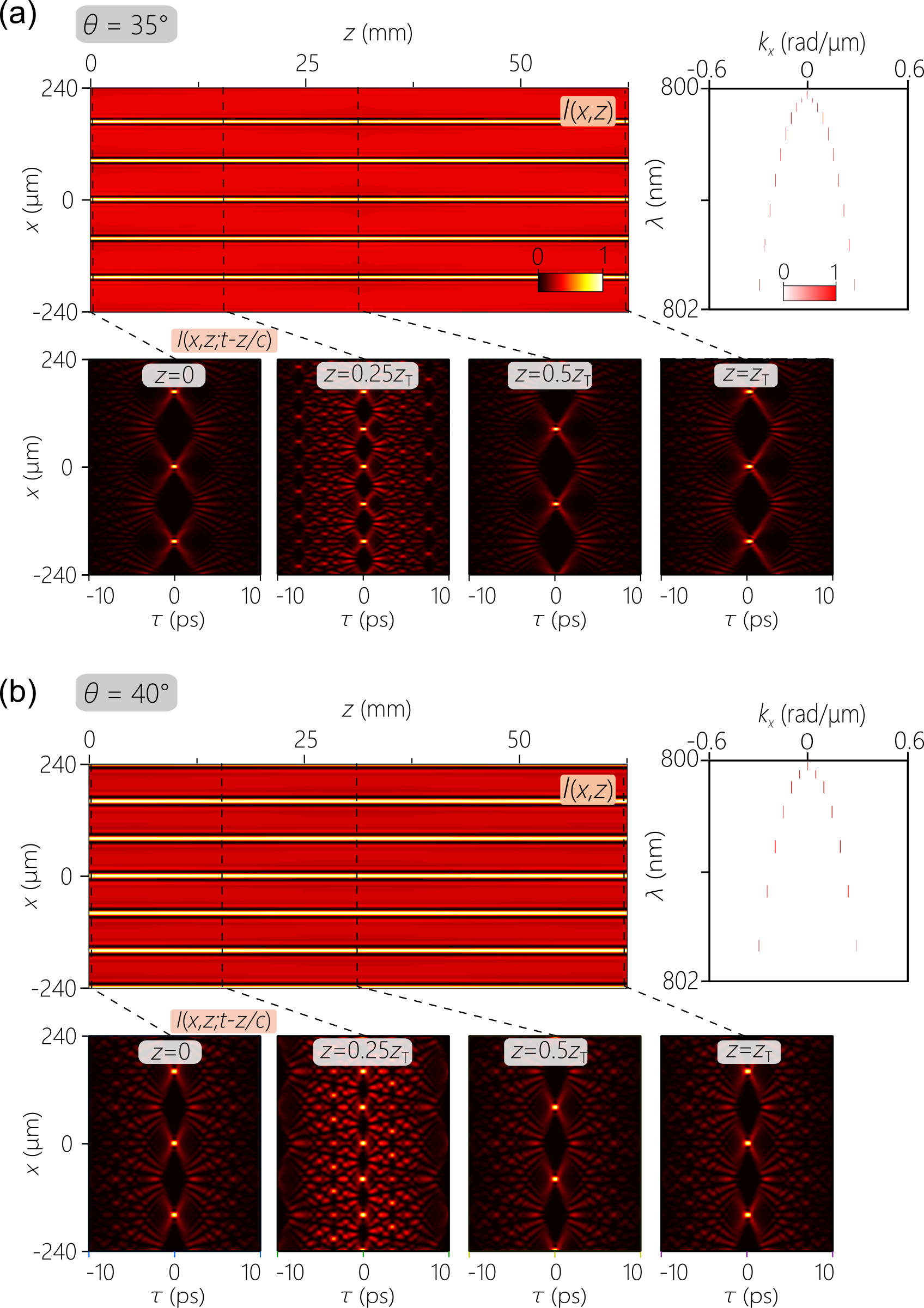}
  \end{center}
  \caption{Simulations of the veiled Talbot ST effect for subluminal ST wave packets with $\lambda_{\mathrm{o}}\!=\!800$~nm, $\Delta\lambda\!=\!2$~nm, and $L\!=\!160$~$\mu$m. (a) The spectral tilt angle is $\theta\!=\!35^{\circ}$. Top row shows the time-averaged intensity $I(x,z)$ over an axial propagation distance of 64~mm, and the discretized spatio-temporal spectral intensity. The lower panels are the calculated spatio-temporal intensity profiles calculated at $z\!=\!0$, 16, 32, and 64~mm. (b) Same as (a) for $\theta\!=\!40^{\circ}$. The transverse spatial period in $I(x,z)$ is $\tfrac{L}{2}\!=\!80$~$\mu$m for both values of $\theta$.}
  \label{Fig:SimulationsTheta1}
\end{figure}

\begin{figure}[t!]
  \begin{center}
  \includegraphics[width=8.6cm]{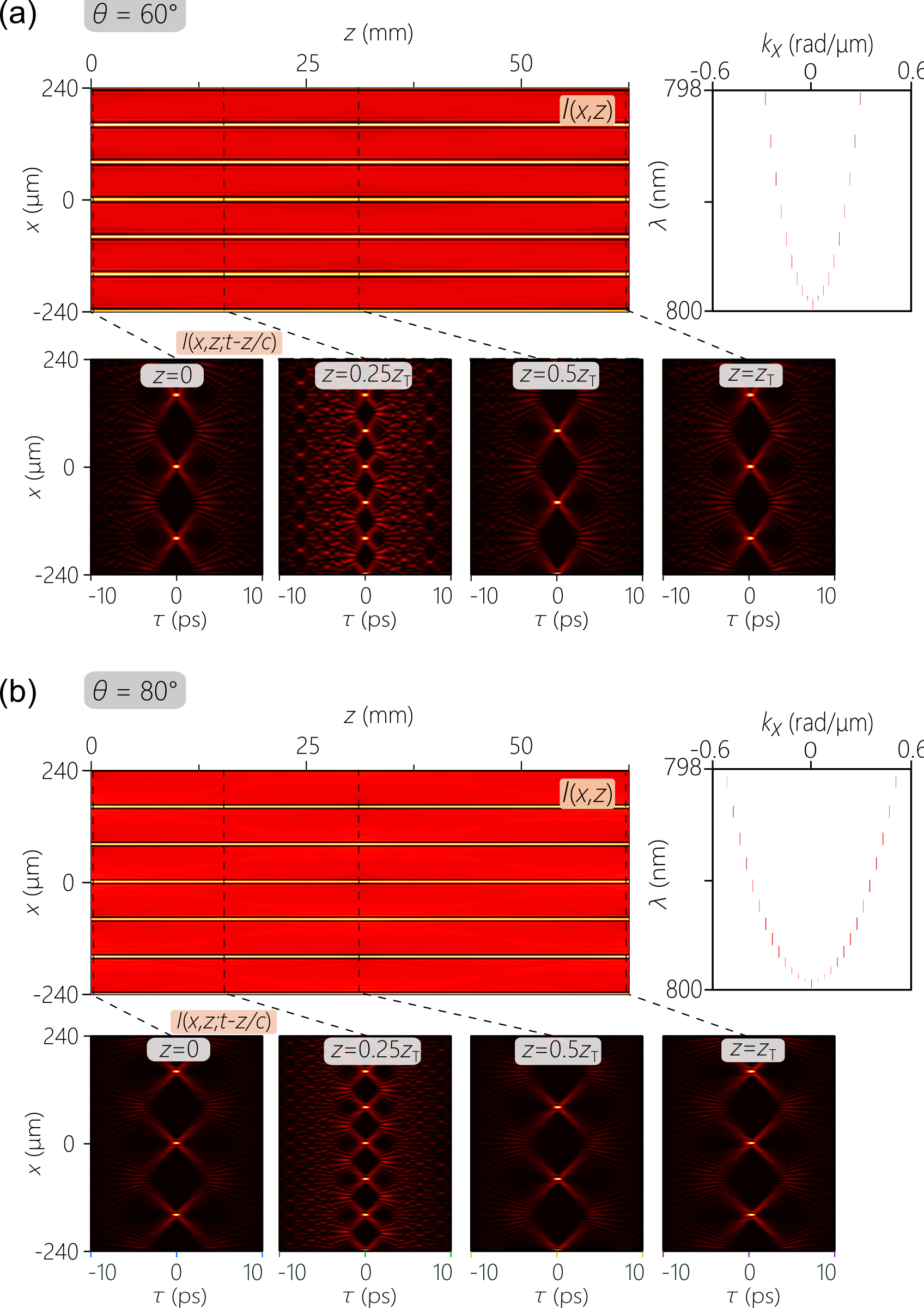}
  \end{center}
  \caption{Simulations of the veiled Talbot ST effect for superluminal ST wave packets with $\lambda_{\mathrm{o}}\!=\!800$~nm, $\Delta\lambda\!=\!2$~nm, and $L\!=\!160$~$\mu$m. (a) The spectral tilt angle is $\theta\!=\!60^{\circ}$. Top row shows the time-averaged intensity $I(x,z)$ over an axial propagation distance of 64~mm, and the discretized spatio-temporal spectral intensity. The lower panels are the calculated spatio-temporal intensity profiles calculated at $z\!=\!0$, 16, 32, and 64~mm. (b) Same as (a) for $\theta\!=\!80^{\circ}$. The transverse spatial period in $I(x,z)$ is $\tfrac{L}{2}\!=\!80$~$\mu$m for both values of $\theta$.}
  \label{Fig:SimulationsTheta2}
\end{figure}

In the experiments reported in the main text, we made use of a spectral tilt angle $\theta\!=\!80^{\circ}$ corresponding to a superluminal group velocity of $\widetilde{v}\!=\!c\tan{80^{\circ}}\!=\!5.67c$. We present here simulations of the veiled ST Talbot effect for other values of the spectral tilt angle spanning the subluminal and superluminal regimes. In \ref{Fig:SimulationsTheta1} we plot simulations for subluminal ST wave packets for $\theta\!=\!35^{\circ}$ ($\widetilde{v}\!=\!0.7c$) and $\theta\!=\!40^{\circ}$ ($\widetilde{v}\!=\!0.84c$); in \ref{Fig:SimulationsTheta2} we plot simulations for superluminal ST wave packets for $\theta\!=\!60^{\circ}$ ($\widetilde{v}\!=\!1.73c$) and $\theta\!=\!80^{\circ}$ ($\widetilde{v}\!=\!5.67c$) for comparison. The spectral uncertainty used in the calculations is $\delta\lambda\!=\!15$~pm.

In all cases, we note the self-revivals at the Talbot planes, the shift by $\tfrac{L}{2}$ between the Talbot planes and the mid-Talbot planes, and the doubling of the transverse period at the quarter-Talbot planes. The only impact of the spectral tilt angle is a change in the temporal scale at each axial plane.

\subsection{Description of the Supporting movies}

Three Supplementary Movies have been provided that offer a complete picture of the evolution of the veiled ST Talbot effect. Each frame in he movie combines two panels. The top panel is corresponds to the time-averaged intensity $I(x,z)$, which shows the diffraction-free behavior of the ST wave packet and a transverse period of $\tfrac{L}{2}$ rather than $L$. A vertical line identifying an axial plane $z$ moves from the left (the source) to the right. The lower panel is the spatio-temporal profile $I(x,z;t)$ corresponding to the axial plane $z$ identified simultaneously in the upper panel. The temporal window over which the profile is shown moves at a group velocity of $c$.

Parameters common to the three movies are the temporal bandwidth $\Delta\lambda\!=\!2$~nm and the wavelength $\lambda_{\mathrm{o}}\!=\!800$~nm, which are selected to match the experimental configuration. The spectral uncertainty used in the calculations is $\delta\lambda\!=\!15$~pm.

\noindent
\textit{\textbf{Supplementary Movie~1:}} This movie shows a continuous simulation of the veiled ST Talbot effect along the axis of a ST wave packet having a periodic transverse profile at $z\!=\!0$. Here we have $L\!=\!100$~$\mu$m and a spectral tilt angle of $\theta\!=\!80^{\circ}$. The simulations follow the center of the wave packet over a full Talbot length of $z_{\mathrm{T}}\!=\!25$~mm.

\noindent
\textit{\textbf{Supplementary Movie~2:}} This movie shows a continuous simulation of the veiled ST Talbot effect along the axis of a ST wave packet similar to that depicted in Supplementary Movie~1 ($L\!=\!100$~$\mu$m), except that the spectral tilt angle has now changed to $\theta\!=\!60^{\circ}$. The simulation follows the center of the pulse over a distance of $z_{\mathrm{T}}\!=\!25$~mm.

\noindent
\textit{\textbf{Supplementary Movie~3:}} This movie shows a continuous simulation of the veiled ST Talbot effect along the axis of a ST wave packet similar to that depicted in Supplementary Movie~1 (spectral tilt angle $\theta\!=\!80^{\circ}$) except that the spatial spectrum is one-sided. That is, only positive spatial frequencies are allowed, thus resulting in a time-averaged spatial profile with no observable spatial features. Here we have $L\!=\!80$~$\mu$m, and the simulations follow the center of the pulse over a distance of $z_{\mathrm{T}}\!=\!16$~mm.

\begin{figure}[t!]
  \begin{center}
  \includegraphics[width=8.6cm]{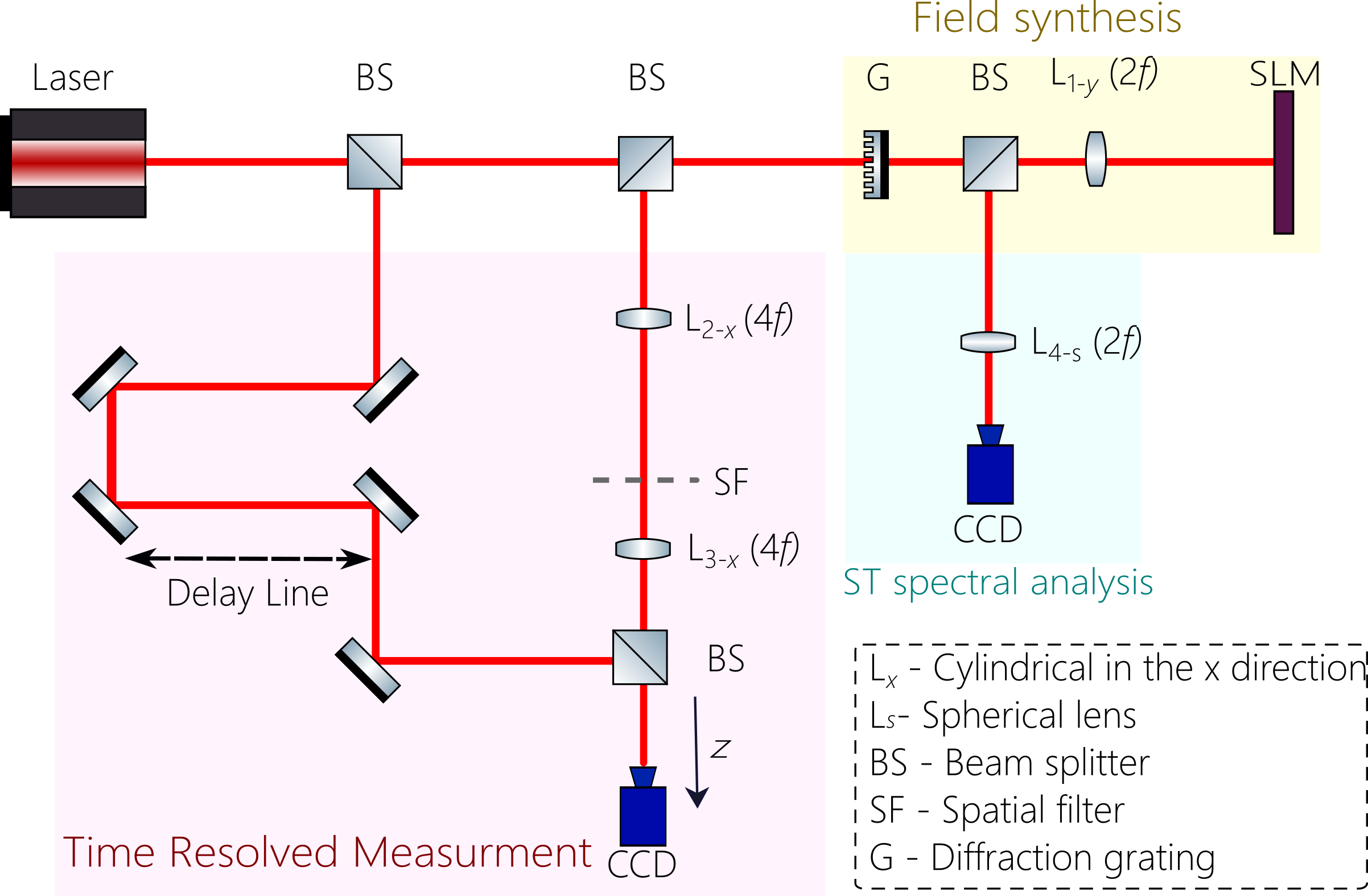}
  \end{center}
  \caption{Schematic depiction of the optical arrangement for the synthesis of ST wave packets. The acronyms for the optical components are all explicated in the legend. The setup consists of three sections highlighted with different background colors: (1) ST field synthesis consisting of a cylindrical lens in a $2f$ configuration with a diffraction grating and SLM; (2) ST spectral analysis, which comprises a single spherical lens and CCD camera; and (3) time-resolved measurements performed via a Mach-Zehnder interferometer that superposes a reference pulse and the synthesized ST wave packet to measure its time-resolved spatial profile.}
  \label{Fig:Setup}
\end{figure}

\section{Experimental configuration}

\subsection{Experimental setup}

The experimental setup for synthesizing ST wave packet is similar to that used in our previous work \cite{Kondakci17NP,Yessenov19OPN}, and combines elements from ultrafast pulse shaping and spatial beam modulation. Plane-wave pulses at a wavelength of $\sim800$~nm from a Ti:sapphire laser (Tsunami Spectra Physics, $\sim100$-fs pulsewidth) are incident on a diffraction grating (Newport 10HG1200-800-1) followed by a cylindrical lens in a $2-f$ configuration; see Fig.~\ref{Fig:Setup}. At the Fourier plane of the lens, where the spectrum of the optical wave is mapped along the horizontal direction ($y$-axis), we impart a two-dimensional phase distribution to the spectrally resolved wave front using a reflective phase-only spatial light modulator (SLM; Hamamatsu X10468-02). In contrast to our previous work, here the SLM imparts a phase pattern that is discretized along the wavelength dimension \cite{Kondakci17NP,Yessenov19OE, Yessenov19unpub}. Such a pattern corresponds to a discrete set of paired spatial and temporal frequencies. The field with the discrete spatio-temporal spectral support domain is retro-reflected and the pulse is reconstituted at the diffraction grating, thereby generating the ST wave packet. 

\begin{figure*}[t!]
  \begin{center}
  \includegraphics[width=12.6cm]{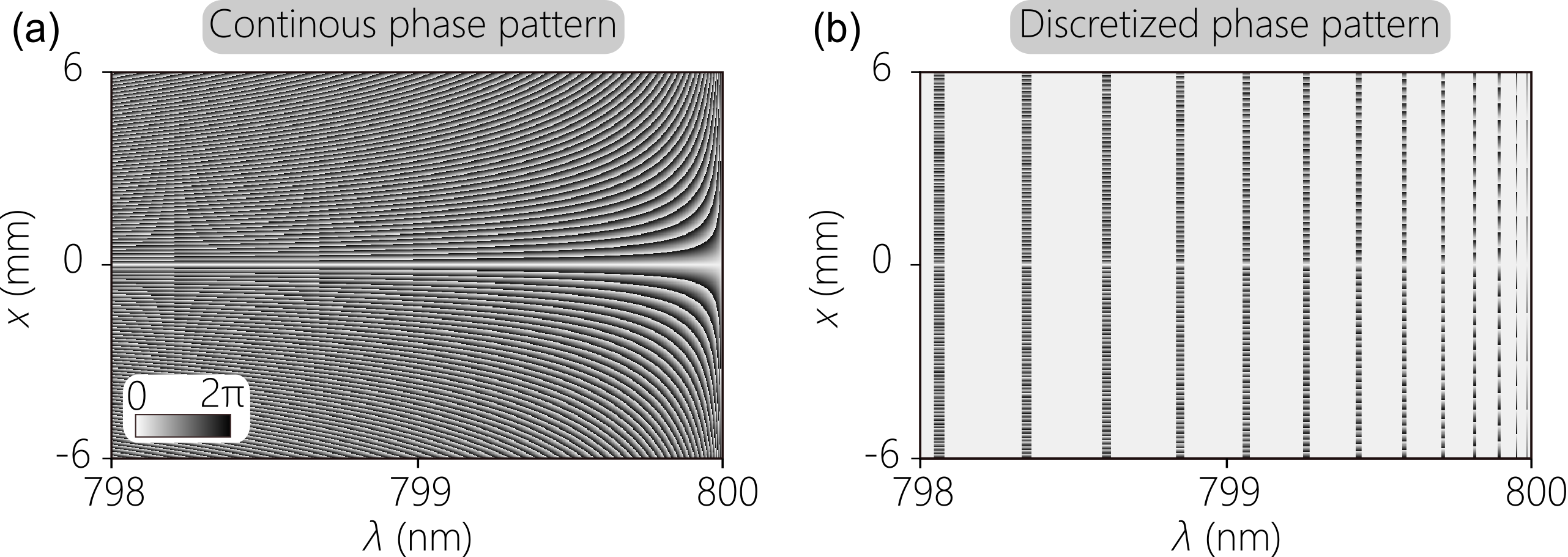}
  \end{center}
  \caption{(a) Two-dimensional phase pattern imparted by the SLM to the incident spectrally resolved wave front. The wavelengths are resolved along the horizontal axis, and the phase in each column is $k_{x}x$ (a phase factor $e^{ik_{x}x}$) with $k_{x}(\Omega)$ selected to realize $\theta\!=\!80^{\circ}$. (b) The phase pattern from (a) discretized along the temporal frequency (wavelength) quadratically $\Omega\rightarrow n^{2}\Omega_{L}$. Consequently, the spatial frequencies in the columns take on discrete values $k_{x}\rightarrow nk_{L}$. In both (a) and (b), $\lambda_{\mathrm{o}}\!=\!800$~nm, $\Delta\lambda\!=\!2$~nm, and $\theta\!=\!80^{\circ}$, and in (b) we have $L\!=\!160$~$\mu$m.}
  \label{Fig:PhasePattern}
\end{figure*}

\subsection{Designing the SLM phase pattern}

In order to associate the desired $k_{x}$ value to each wavelength, a diffraction grating is used to spread the temporal spectrum in space, and a cylindrical lens collimates the separated wavelengths along the $y$-axis as shown in Fig.~\ref{Fig:PhasePattern}. Each wavelength occupies a column on the SLM ($x$-axis) where a linear phase pattern of the form $k_{x}x$ (modulo $2\pi$) is implemented, which corresponds to assigning the spatial frequency $k_{x}$ to this wavelength. An example corresponding to $\theta\!=\!80^{\circ}$ is shown in Fig.~\ref{Fig:PhasePattern}(a) for a continuous temporal spectrum. In the case of a discrete spectrum, $k_{x}\rightarrow n\tfrac{2\pi}{L}$ and $\Omega\rightarrow n^{2}\Omega_{L}$ ($n$ integer), which is achieved by activating only specific columns on the SLM (corresponding to the desired frequencies $n^{2}\Omega_{L}$) and along each column we implement the phase associated with the corresponding spatial frequencies $nk_{L}$. To provide sufficient optical signal, we select a finite width for each column of the phase pattern along $y$ for each wavelength. The phase pattern along $x$ is maintained the same across the width of each column (fixed spatial frequency, $nk_{L}$). The width of the columns increases the spectral uncertainty $\delta\omega$, thereby reducing the temporal width of the pilot envelope as described above \cite{Yessenov19OE}.

\begin{figure*}[t!]
  \begin{center}
  \includegraphics[width=13.6cm]{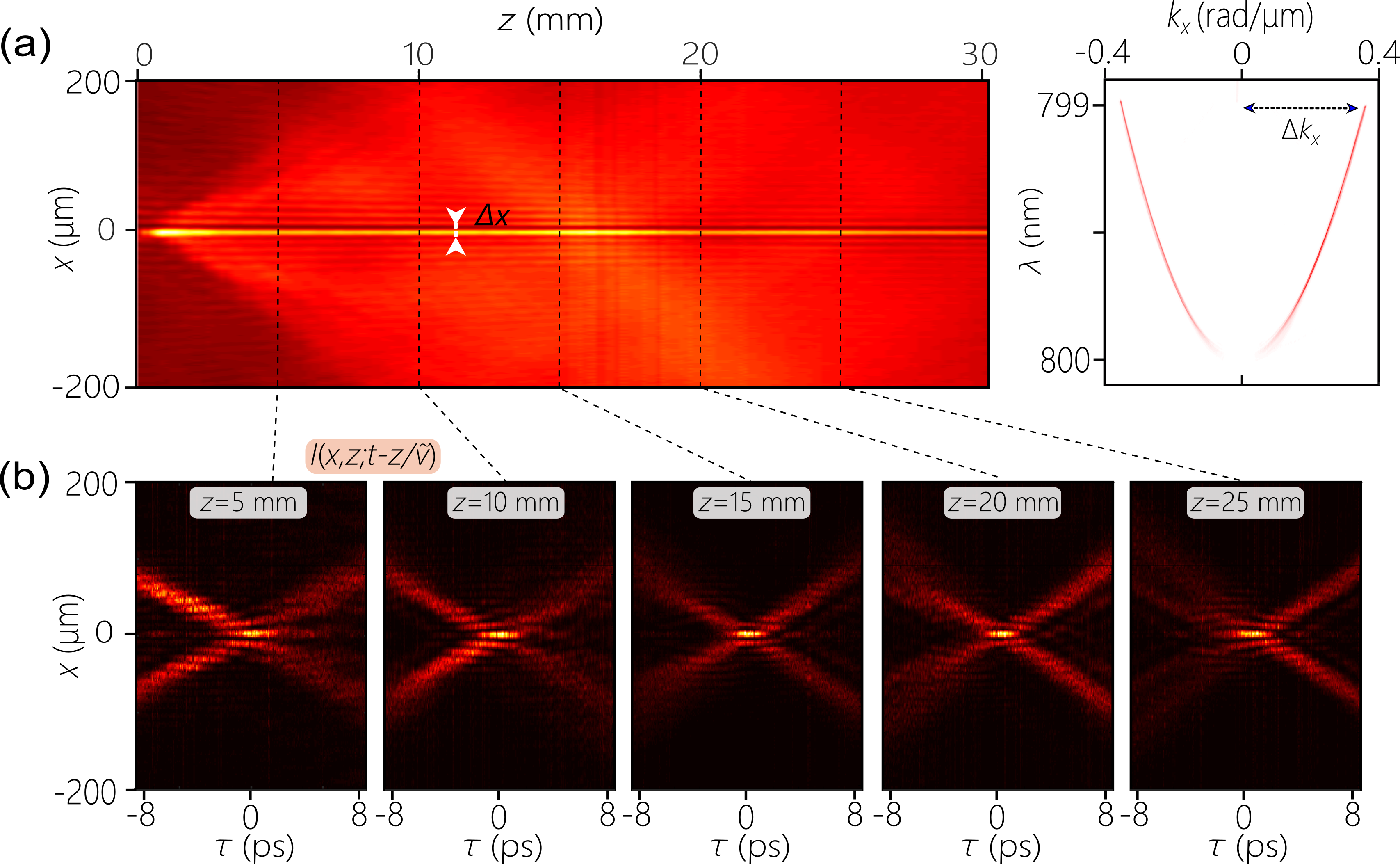}
  \end{center}
  \caption{(a) Measured time-averaged intensity $I(x,z)$ for a ST wave packet having the same parameters as Fig.~2 in the main text except that the spatio-temporal spectrum is continuous rather than discrete. The measured spatio-temporal spectral intensity is plotted on the right. (b) Measured spatio-temporal profiles $I(x,z;\tau)$ in a moving reference frame moving at a group velocity $\widetilde{v}$ and \textit{not} $c$, obtained at different axial planes $z$ (taken at 5-mm axial intervals). The spatio-temporal profile does not change axially in contrast to the periodic ST wave packets resulting from spectral discretization.}
  \label{Fig:ContinuousST}
\end{figure*}

\subsection{Characterization}

The synthesized ST field is characterized in three different domains.

\begin{enumerate}
    \item \textit{\textbf{Fourier-space analysis}}. The spatio-temporal spectrum is characterized in Fourier space $(k_{x},\lambda)$ to ensure that the desired spatio-temporal spectral support domain is indeed implemented by the SLM. By sampling the spectrally resolved field of the retro-reflected field from the SLM  and implementing a spatial Fourier transform via the spherical lens in a $2-f$ configuration, we resolve the field in $(k_{x},\lambda)$ space at the focal plane of the lens, where it is captured by a CCD camera (CCD$_1$; TheImagingSource, DMK 33UX178); see Fig.~\ref{Fig:Setup}. Using this approach, we obtained the data in the main text in Fig.~2(a,c) and Fig.~3(a). 
    \item \textit{\textbf{Spatial intensity profiling in physical space}}. To record the time-averaged intensity $I(x,z)$, we scan a CCD camera (CCD$_2$; TheImagingSource, DMK 27BUP027) along the propagation axis $z$; see Fig. \ref{Fig:Setup}. Using this approach, we obtained the data in the main text in Fig.~2(a,c) and Fig.~3(a). 
    \item \textit{\textbf{Time-resolved spatial profile}}. Using a Mach-Zehnder interferometer in which with a short reference laser pulse (100-fs pulsewidth) from the Ti:sapphire laser directed to a reference arm containing an optical delay line, and the ST wave packet synthesized in the second arm, interference of the two wave packets occurring when they overlap in space and time reveals high-visibility spatially resolved fringes. From this we obtain the time-resolved intensity $I(x,z;\tau)$ at locations of interest along $z$-axis. Using this approach, we obtained the data in the main text in Fig.~2(b,d) and Fig.~3(b).
\end{enumerate}

In the main text (Fig.~2 and Fig.~3), we presented measurements for ST wave packets having a periodic transverse profile that result from spectral discretization. The measurements in the text demonstrated the veiled ST Talbot effect. Specifically, the measured spatio-temporal profiles $I(x,z;\tau-\tfrac{z}{c})$ was measured in a reference frame moving at a group velocity of $c$, and the profiles changed along the propagation axis $z$. We present here for comparison measurements of the same ST wave packet in absence of spectral discretization. We make use of the same parameters for the ST wave packet in Fig.~2 in the main text: $\Delta\lambda\!=\!2$~nm, $\lambda_{\mathrm{o}}\!=\!800$~nm, and $\theta\!=\!80^{\circ}$. The measurements are plotted in \ref{Fig:ContinuousST}. The transverse profile of the time-averaged intensity $I(x,z)$ (\ref{Fig:ContinuousST}A) is no longer periodic. Crucially, the spatio-temporal profiles (\ref{Fig:ContinuousST}B) are invariant along $z$ in contradistinction to the axially varying results when the spectrum is discretized and the transverse profile is periodic. Furthermore, the measurements are carried out in a reference frame moving at a group velocity of $\widetilde{v}\!=\!c\tan{\theta}$ rather than $c$.

\bibliography{bibleog}